\documentclass[useAMS,usenatbib]{mn2e}
\usepackage[dvips]{graphicx}
\usepackage{amssymb}
\def\aj{AJ}%
\def\apj{ApJ}%
\def\apjl{ApJ}%
\def\apjs{ApJS}%
\def\aap{A\&A}%
\def\aaps{A\&AS}%
\def\mnras{MNRAS}%
\def\pasp{PASP}%
\def\pasj{PASJ}%
\def\nat{Nature}%
\title[Intracluster Light at $z\sim0.25$ from SDSS image stacking]{
Intergalactic stars in $z\sim0.25$ galaxy clusters:
systematic properties from stacking of Sloan Digital Sky Survey imaging data}
\author[S. Zibetti et al.]{Stefano Zibetti$^{1,2}$\thanks{E-mail:
szibetti@mpe.mpg.de} and Simon D. M. White$^{1}$\\
$^{1}$Max-Planck-Institut f\"ur Astrophysik, Karl-Schwarzschild-Str. 1, 
D-85748 Garching bei M\"unchen, Germany\\
$^{2}$Max-Planck-Institut f\"ur Extraterrestrische Physik, Giessenbachstrasse, 
D-85748 Garching bei M\"unchen, Germany
\newauthor
Donald P. Schneider$^{3}$, Jon Brinkmann$^{4}$\\
$^{3}$Department of Astronomy and Astrophysics, Pennsylvania State University, 
University Park, PA 16802, USA\\
$^{4}$Apache Point Observatory, PO Box 59, Sunspot, NM 88349, USA}
\begin{document}

\date{Accepted . Received ; in original form }

\pagerange{\pageref{firstpage}--\pageref{lastpage}} \pubyear{2005}

\maketitle

  \label{firstpage}

\begin{abstract}
We analyse the spatial distribution and colour of the intracluster light (ICL) in 683 clusters of
galaxies between $z=0.2$ and 0.3, selected from $\sim 1500$ deg$^2$ of the first data release of the
Sloan Digital Sky Survey (SDSS-DR1). Surface photometry in the $g$, $r$ and $i$ bands is conducted
on stacked images of the clusters, after rescaling them to the same metric size and masking
out resolved sources. We are able to trace the average surface brightness
profile of the ICL out to 700 kpc, where it is less than $10^{-4}$ of the mean surface brightness 
of the dark night sky. 
The ICL appears as a clear surface brightness excess with respect to an inner
$R^{1/4}$ profile which characterises the mean profile of the brightest cluster galaxies (BCG).
The surface brightness of the ICL ranges from 27.5 mag arcsec$^{-2}$ at 100 kpc to $\sim 32$ mag 
arcsec$^{-2}$ at 700 kpc in the observed $r$-band. This corresponds SB in the range 26.5 to 31
in the rest-frame $g$-band.
We find that, on average, the ICL contributes only a small fraction 
of the total optical emission in a cluster. Within a fixed metric aperture
of 500 kpc, this fraction is $10.9\pm5.0$ per cent for our clusters. 
A further $21.9\pm3.0$ per cent is contributed on average by the BCG itself. The radial 
distribution of the ICL is more centrally concentrated than that of the cluster galaxies, but
the colours of the two components are identical within the statistical uncertainties.
In the mean the ICL is aligned with and {\it more flattened} than the BCG itself.
This alignment is substantially stronger than that of the cluster light as a whole.
We find the surface brightness of the ICL to correlate both with BCG luminosity and with
cluster richness, while
the fraction of the total light in the ICL is almost independent of these quantities.
These results support the idea that the ICL is produced by stripping and disruption of galaxies
as they pass through the central regions of clusters.
Our measurements of the diffuse light also constrain the faint-end slope of the cluster luminosity
function. Slopes $\alpha<-1.35$ would imply more light from undetected galaxies than is observed
in the diffuse component.
\end{abstract}

\begin{keywords}
galaxies: clusters: general; galaxies: elliptical and lenticular, cD; diffuse radiation; 
galaxies: interactions; galaxies: evolution; galaxies: formation
\end{keywords}

\section{Introduction}
Firstly proposed by \cite{zwicky51}, the presence of a diffuse population of intergalactic stars
in galaxy clusters is now a well established observational fact. After pioneering work
in the 1970s based on photographic plates \citep[e.g.][]{welch_sastry_71} and photoelectric 
detectors \citep{melnick_white_hoessel77}, CCD detectors have 
made it possible to conduct deep surveys in nearby galaxy clusters and to detect unambiguously
the intracluster light (ICL) 
\citep[e.g.][]{bernstein+95,gonzalez+00,feldmeier+02,feldmeier+04,gonzalez+04}.
Parallel searches for resolved intracluster stars \citep{durrell+02} and planetary nebulae
\citep[e.g.][]{arnaboldi+96,feldmeier+04c} have confirmed the presence of a population of stars which are
not dynamically  bound to any individual galaxy, but orbit freely in the cluster potential.
The ICL contributes a substantial fraction of the optical emission in a cluster.
Estimates range from 
approximately 50 per cent in the core of the Coma cluster (\citealt{bernstein+95}, although
an upper limit of 25 per cent was found over a larger region by \citealt{melnick_white_hoessel77}),
to 10--20 per cent in less
massive clusters \citep{feldmeier+04}. Arclets and other
morphologically similar low surface brightness features have been identified by several authors in 
the Coma
and Centaurus clusters \citep{gregg_west98,trentham_mobasher98,calcaneo+00}, suggesting that
at least part of the
ICL is contributed by dynamically young tidal features. This supports the commonly accepted idea 
that this stellar population is made up of disrupted dwarf galaxies and of stars stripped from more 
massive galaxies.\\

We are still far from a complete understanding of the physical mechanisms that produce
the ICL.
Several 
mechanisms can act to remove stars from individual galaxies and to fling them into intergalactic
space. The relative importance and effectiveness of these mechanisms can vary during the 
evolutionary history of
a cluster and from place to place within the cluster. Tides generated by the
cluster potential \citep{merritt84} and repeated high speed encounters between galaxies 
\citep{richstone76} are the dominant stripping mechanisms in a fixed cluster potential,
as demonstrated graphically by the simulations of \cite{moore+96}. However,
when the evolution of the cluster and the presence of substructures is taken into account, two
other mechanisms become relevant \citep{gnedin03}.
Preprocessing of the ICL occurs during low velocity encounters between galaxies 
within the groups which eventually merge into the cluster, and galaxies are dynamically heated 
by encounters with substructures. As stressed by \cite{mihos04}, the tidal tails and the heated 
structures preprocessed within groups are subsequently easily removed by the cluster potential.

In order to encompass all these processes in a cosmologically motivated framework, many groups 
in recent years have addressed the ICL problem using high resolution N-body simulations 
\citep{napolitano+03}, some including smoothed
particle hydrodynamics (SPH) to take gas processes into account
\citep{murante+04,willman+04,sommerlarsen+04}. Although many issues still have to be clarified and
agreement between the different models is far from complete, these simulations show how the ICL may
be produced over the entire history of the clusters from continuous stripping of
member galaxies and through contributions from merging groups.
Unfortunately the degree to which they correctly represent the internal structure of cluster galaxies
and their dark matter haloes is too uncertain for their prediction for the amount of ICL to be 
reliable.\\

Although the number of observations of the ICL in individual clusters has increased rapidly in 
recent years, we still lack a large sample that allows generalisation of the
properties of the ICL and an understanding of how they depend on global cluster properties, 
particularly on richness and on the
luminosity of the first ranked galaxy. Given the very low surface brightness of the ICL,
typically less than 0.1 per cent of that of the night sky, observations
are extremely challenging. Not only are long exposure times required in order to obtain acceptable
signal-to-noise (S/N) ratios, but many subtle instrumental effects, such as flat fielding inhomogeneities
and scattered light within the camera, must be kept under tight control.

An alternative approach, which exploits the wealth of imaging data
made available by the Sloan Digital Sky Survey \citep[SDSS, ][]{SDSS}, has been proposed and 
successfully applied to the
study of stellar haloes around galaxies by \cite{z04}. By stacking several hundreds of images, 
mean surface 
brightnesses of the order of $\mu_r$=29--30 mag arcsec$^{-2}$ can be reliably measured. In fact,
not only is the S/N enhanced, but also inhomogeneities in the background signal and in flat fielding
are averaged out. A further advantage comes from the fact that the SDSS images are obtained in
drift-scan mode \citep{gunn_etal98}. As opposed to the `staring' mode, in which the intensity of 
each pixel in the image is measured by the corresponding pixel on the CCD array, in drift-scan mode
the signal is integrated over an entire column of the CCD while the target drifts in the field of
view. Therefore, sensitivity variations can occur only in one dimension (i.e. perpendicular
to the drift direction) instead of two; this strongly reduces the flat-field inhomogeneities
in the frames.

The stacking analysis is statistical in its nature, providing mean results for large samples
of galaxy clusters, which can be compared in principle to similar properties derived from 
cosmological
simulations. An appropriate choice of subsamples makes it possible to study the influence of
different parameters on the properties of the ICL. Although high statistical significance is the
main advantage of the stacking method, individual features (tidal streams and
arclets, for instance) and real cluster-to-cluster variations are lost in the averaging.
The stacking method is therefore complementary to imaging
of individual clusters, from which detailed information about small scale structures and
stochastic phenomena can be derived.\\

In this paper we present an analysis of the stacking of 683 clusters imaged in the $g$, $r$, and $i$
bands in the SDSS. They were selected over $\sim 1500$ deg$^2$  between $z=0.2$ and 0.3, using
the maxBCG method (Annis et al., in preparation). Details on the sample selection and 
on sample properties are given
in Sec. \ref{sample}. The image processing and the stacking technique are described
in Sec. \ref{proc_stack}. In Sec. \ref{photo_analysis} we describe how the relevant photometric
quantities are derived. We present the results of our analysis in Sec. \ref{results}.
Possible sources of systematic uncertainties on the derived quantities are discussed in
Sec. \ref{systematics}. Our results are
compared to other extant observations and model predictions in Sec. \ref{discussion}, and
some possible implications for theories of the formation of the ICL during cluster evolution
are presented. Conclusions and future perspectives are outlined in Sec. \ref{conclusions}.\\
Throughout the paper we adopt the ``concordance'' cosmology, $H_0=70$ km sec$^{-1}$ Mpc$^{-1}$, 
$\Omega_0$=1, $\Omega_\Lambda$=0.7.

\section{The sample}\label{sample}
The imaging data utilised in this work are derived from the SDSS \citep{SDSS,EDR,DR1,DR2}.
The SDSS is imaging about a quarter of the sky in the $u$, $g$, $r$, $i$, and $z$ bands, 
with a $\sim$54 sec exposure in drift scan mode at the dedicated 2.5 m Apache Point Observatory 
telescope
\citep{fukugita_etal96,gunn_etal98,hogg_etal01,smith_etal02,pier_etal03}, reaching 
$\sim25~\mathrm {mag~arcsec^{-2}}$ at $S/N\sim1$ for a single pixel in the $r$--band. 
The SDSS spectroscopic galaxy samples \citep{blanton_tiling03} consist of all galaxies brighter 
than $r=$17.77 mag \citep{strauss_etal02}
and of a sample of Luminous Red Galaxies \citep{eisenstein+01} extending at $r<19.2$ mag.\\
To reach surface brightnesses as low as 29--30$~\mathrm {mag~arcsec^{-2}}$ stacking
of several hundreds of images is required.
We have focused our sample selection on clusters in the redshift range 0.2--0.3 in order
to satisfy the requirement of homogenous imaging coverage within each cluster. Along the scan
direction the main limitation is given by the sky background fluctuations, whereas
in the perpendicular direction the limit is given by the width of the SDSS camera columns
\citep{gunn_etal98}, which corresponds to $\sim13.5$ arcmin. For practical convenience we use 
only the SDSS fpC frame ($13.5\times 9.8$ arcmin$^2$) in which the cluster centre is located and 
require that
a significant fraction of ``background'' beyond 1 Mpc projected distance from the cluster 
centre be included. Given that 1 Mpc=5.05 arcmin at $z=0.2$, this turns out to be a good lower 
redshift limit. 
On the other hand, we prefer to avoid extending the sample to much higher redshift, both
because cosmological dimming acts to reduce the apparent surface brightness by $(1+z)^4$, and
because resolving individual galaxies in the clusters becomes increasingly difficult. Moreover,
K-corrections for band shifting would have to be taken into account in order to interpret a 
stack of objects in a wide range of $z$. Since we do not know {\it a priori} the spectral 
energy
distribution (SED) of the ICL, this would add considerable uncertainty to results. The 
4000\AA--break is probably the main feature in the SED of the ICL, so we have chosen $z=0.3$ 
as upper limit; the break is then almost homogeneously bracketed by the $g-r$ colour over the 
whole sample.\\

Cluster identifications over an area of $\sim 1500$ deg$^2$ in the SDSS DR1 \citep{DR1} have been
kindly provided by J. Annis, based on the maxBCG method 
\citep[see Annis et al., in preparation,][ for details]{bahcall+03}. This method is based on the 
fact that (i) the BCGs lie in a narrow region of the $(g-r)$-$(r-i)$-$M_i$ space, and (ii) the 
early type galaxies in a cluster define a ridge-line in the colour-magnitude diagram. The likelihood 
of a galaxy being the BCG of a cluster is calculated for a grid of different redshifts taking 
into account the ``distance'' of the galaxy from the predicted BCG locus and the number of galaxies
$N_{\mathrm{gal}}$ within 1 $h^{-1}$ Mpc which lie less than 2 $\sigma$ away from the early-type
colour-magnitude relation ($\sigma$ being the average scatter of the relation). 
The redshift which maximises the likelihood is taken as the fiducial redshift of the cluster;
only clusters whose probability is greater than a certain threshold are considered.
In addition to the identification of the BCG and the photometric redshift of the cluster, the maxBCG
method produces the number of red-sequence galaxies within 1 $h^{-1}$ Mpc, $N_{\mathrm{gal}}$,
their total luminosity, $L_{\mathrm{RG}}$, and the number of red-sequence galaxies within 0.33 
$h^{-1}$ Mpc, $N_{\mathrm{gal,3}}$. Based on the analysis conducted by \cite{hansen+02} and 
Hansen et al. (in preparation) on the galaxy count overdensity around 12830 BCGs in the redshift 
interval $0.07\leq z < 0.3$, an estimate of $R_{200}$\footnote{$R_{200}$ is the radius that
encloses an average mass density which is 200 times the density of the background.} is given using
the empirical relation found 
by these authors between $N_{\mathrm{gal}}$ and $R_{200}$.\\

From the maxBCG catalogue we have selected all the clusters with: 
$0.15\leq z_{\mathrm{maxBCG}}\leq 0.35$; $N_{\mathrm{gal}}\geq15$; $N_{\mathrm{gal,3}}\geq5$.
This preliminary selection uses a broader redshift range to include those BCGs whose spectroscopic
$z$ is within the 0.2--0.3 interval, although the maxBCG $z$ is not (see below).
The constraints on the number of galaxies within
1 and 0.33 $h^{-1}$ Mpc should ensure that we select clusters with richness similar to those listed in 
the Abell catalogue \citep{abell58,abell_cat}.
The positions of the BCG that passed this first
selection has been matched to objects in the SDSS spectroscopic database. Whenever available,
the spectroscopic redshift of the BCG has been assigned as the fiducial redshift of the entire 
cluster (see Annis et al., in preparation, for details on the precision achieved by the maxBCG 
in the redshift determinations).
All selected clusters with $0.195<z<0.3$ have been inspected using RGB composite images ($g-r-i$) 
in order:
{\it i)} to exclude the images affected by evident defects, such as strong background gradients,
and scattered light from very bright stars; {\it ii)} to check that the selected BCG is actually the
brightest member (the maxBCG algorithm sometimes selects the second or third ranked member),
and, if not, to assign the position of the new BCG;
{\it iii)} to exclude candidate clusters where no clear enhancement of the galaxy number density 
is visible toward the centre. This final step prunes the sample of roughly
10 per cent of poor cluster candidates, with a very low spatial concentration. These are likely just
chance superposition of galaxies, rather than physically bound associations.
A total of 683 clusters satisfy all these requirements and constitute our main sample.

\subsection{Sample properties}\label{subsamples}
\begin{figure}
\includegraphics[width=8truecm]{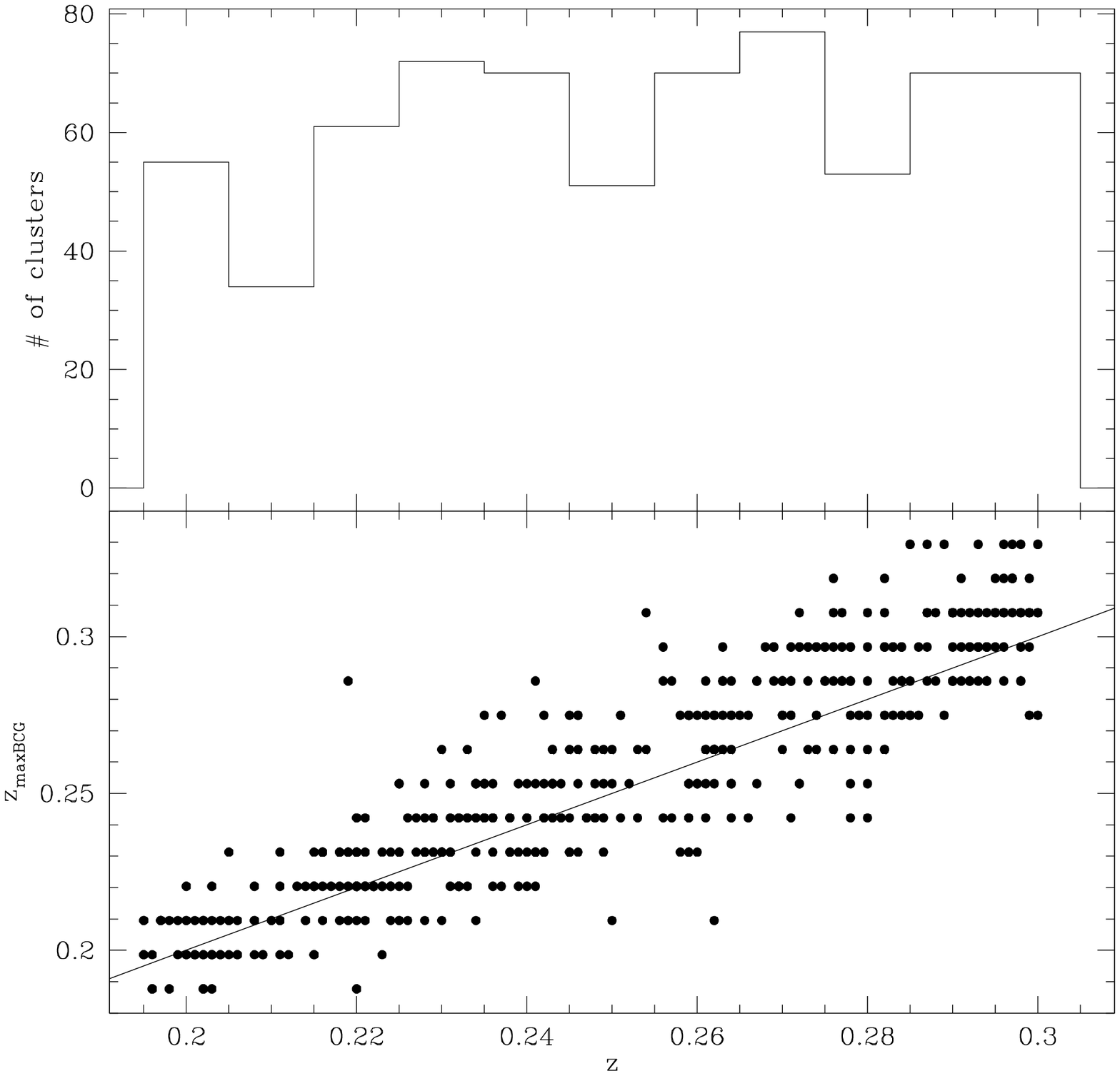}\caption{The redshift distribution of
the galaxy clusters in our main sample (upper panel). In the bottom panel we show the comparison 
between the spectroscopic (abscissa) and the photometric maxBCG redshifts (ordinates) for the
464 clusters whose BCG has been spectroscopically observed. The $z=z_{\mathrm{maxBCG}}$ relation
is shown by the straight line.}\label{redshift}
\end{figure}

The distribution in redshift of the main sample is almost uniform between 0.2 and 0.3, as
shown in the top panel of Fig. \ref{redshift}, and
we will therefore use 0.25 as the reference redshift of the sample.
In the bottom panel of the same figure we plot the spectroscopic (in abscissa) and the photometric
maxBCG redshifts (in ordinates) for the 464 clusters whose BCG has been spectroscopically observed.
The typical error of the photometric redshift is 0.015. Note that 49 of these clusters would have
been excluded based on the photometric redshift, as located beyond $z=0.3$. Considering that among 
the 219 clusters, for which no spectroscopic redshift is available, only 69 are at 
$z_{\mathrm est}>0.28$, we conclude that fewer than 15-20 clusters of $z$ significantly larger than 
0.3 contaminate our sample.

\begin{figure*}
\includegraphics[width=\textwidth]{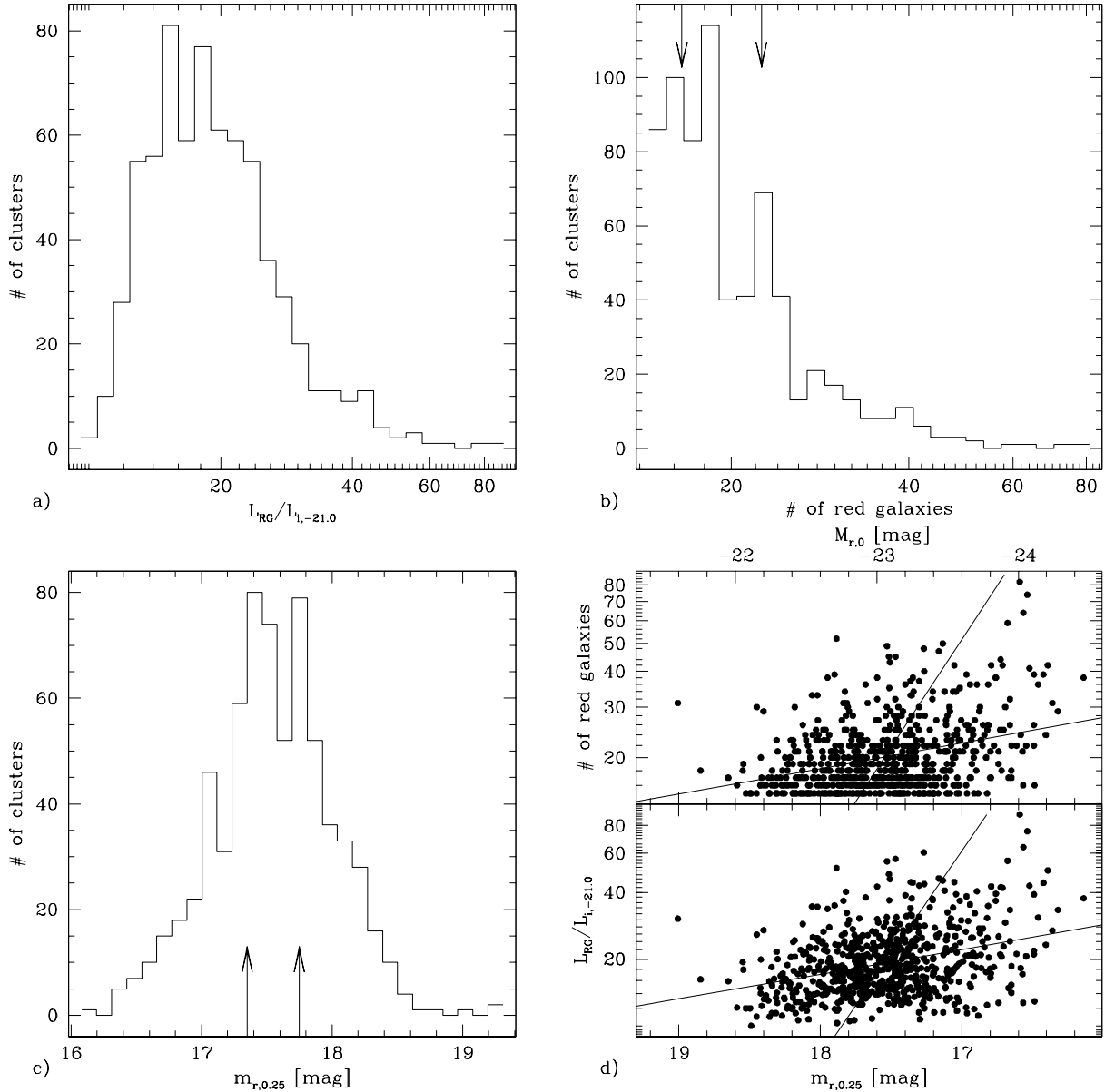}\caption{Properties of the main sample of
galaxy clusters. a) distribution of the total luminosity of the red sequence galaxies. The luminosity
is expressed in units of the luminosity corresponding to $M_i=-21.0$.
b) distribution of the number of red sequence galaxies. Arrows indicate the upper and lower limits
for the ``poor'' and ``rich'' subsamples respectively. c) Distribution of BCG luminosities, in units
of $r$--mag in the observer frame for $z=0.25$. Arrows indicate the upper and lower limits
for the ``luminous-BCG'' and ``faint-BCG'' subsamples respectively. d) Correlations of the BCG luminosity with the 
luminosity of the red sequence galaxies (bottom) and with the number of red sequence galaxies (top).
Solid lines show the direct (y vs. x) and inverse (x vs. y) regression lines from the least squares
linear fitting to the points. The scale on the upper axis reports the rest-frame $r$-band absolute
magnitudes, using the K-corrections calculated for a 13 Gyr old, solar metallicity SSP
with the code of Bruzual \& Charlot (2003).}\label{sample_plot}
\end{figure*}

The distributions of the
other fundamental properties derived from the maxBCG analysis and from the photometry of the BCG
are reported in Fig. \ref{sample_plot}. In the first three panels (a, b and c), we show
histograms of the number of clusters as a function of the luminosity of the red sequence galaxies,
of the number of red galaxies, and of the luminosity of the BCG, respectively. The total luminosity
and the number of red galaxies are in principle equivalent proxies for the richness of clusters.
However, due to the small number of galaxies in the poorest clusters, the luminosity provides
a smoother distribution at the poor end. The three distributions are peaked around the average
values, with roughly 50 per cent of the sample sharing very similar properties, namely
16--24 red galaxies or $L_{\mathrm{RG}}=$15--25$~L_{i,-21.0}$,\footnote{$L_{i,-21.0}$ is the 
luminosity corresponding to $-21.0$ absolute $i$--band mag in the rest-frame.} and $m_{r,0.25}$(BCG)
between 17.2 and 18.
The well known correlation between richness and luminosity of the BCG is present in our
sample as visible in panel d). Nevertheless, the scatter is conspicuous and different BCG 
luminosities can correspond to very different richness.
\\
Unfortunately the redshift range of our sample makes its overlap with the Abell cluster
catalogue quite small: although 130 clusters catalogued by \cite{abell_cat} as distance class 6
are included in the area of sky covered by our sample, only 43 match the position of
our clusters within 6 arcmin, and have Abell richness ranging from 0 to 4. 
Most (95 per cent) of the remaining 87 are excluded because of 
their low redshift, and just 5 per cent are rejected because of defects in the imaging data.\\

\begin{figure*}
\centerline{
\includegraphics[width=8.5truecm]{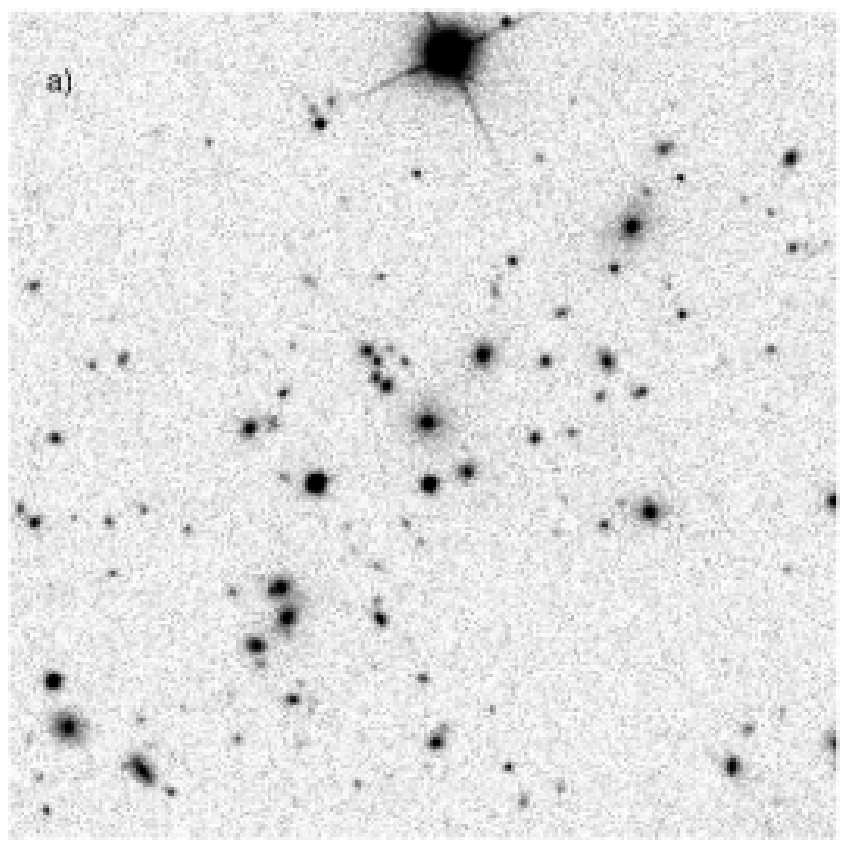}
\includegraphics[width=8.5truecm]{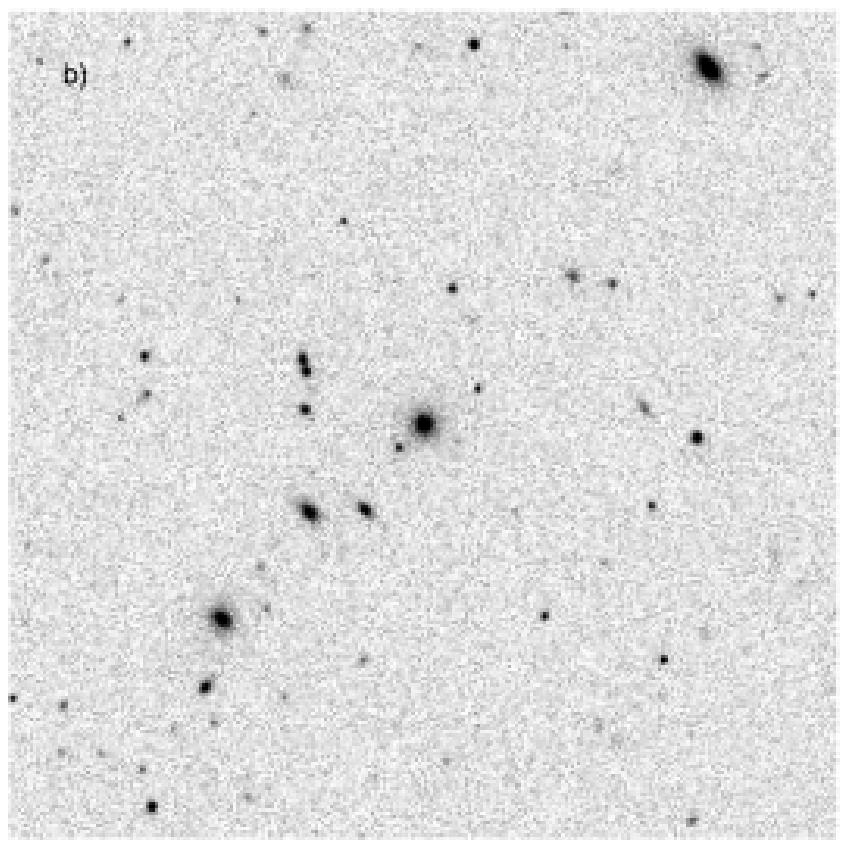}
}
\centerline{
\includegraphics[width=8.5truecm]{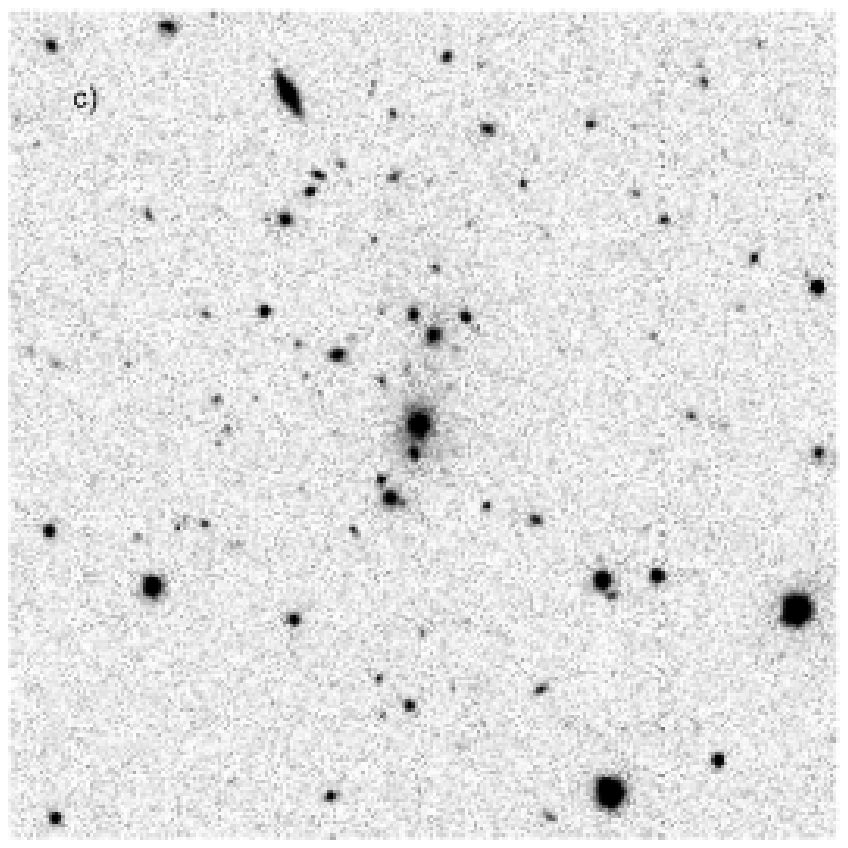}
\includegraphics[width=8.5truecm]{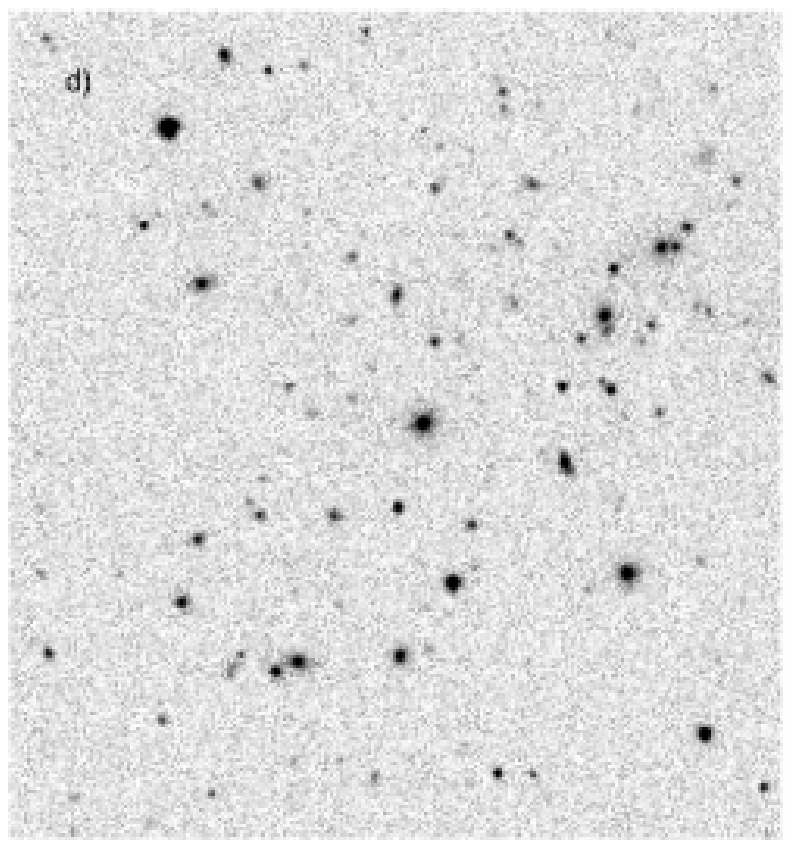}
}\caption{$r$--band images of four typical clusters in the ``rich'', ``poor'', ``luminous-BCG'' 
and ``faint-BCG'' subsamples (panels a, b, c and d respectively). Each frame is $500\times 500$
pixel$^2$, corresponding to 3.3 arcmin side, $\sim0.8 h_{70}^{-1}$ projected Mpc.}\label{examples}
\end{figure*}

In Section \ref{prop_dep} we will analyse the properties of the ICL in different cluster subsamples.
In particular we will refer to ``poor'' (``P'') and ``rich'' (``R'') clusters as those having less 
than 17 red galaxies and more than 22,\footnote{Note that of the 178 clusters that make up the
`rich' subsample, 31 are catalogued by \cite{abell_cat}, corresponding to 17.4 per cent.} 
respectively; ``luminous-BCG'' (``L'') and ``faint-BCG'' (``F'')
clusters are classified according to the luminosity of the BCG, brighter than $m_{r,0.25}=17.35$ or 
fainter than 17.75.
These boundaries are marked with arrows in panels b and c of Fig. \ref{sample_plot}.
In order to illustrate the differences, we show in Fig. \ref{examples} the $r$--band
images of four typical clusters in the ``rich'', ``poor'', ``luminous-BCG'' and ``faint-BCG'' subsamples
(panels a, b, c and d respectively). It is particularly instructive to see how richness and
BCG luminosity do not always correspond, despite their general correlation. Although
the BCG of the ``rich'' cluster is more luminous than the BCG of the ``faint-BCG'' one, it is not 
significantly brighter than the BCG of the ``poor'' cluster. Vice versa, although the ``luminous-BCG'' cluster is 
richer than the ``poor'' one, it is not significantly richer than the ``faint-BCG'' cluster.

\section{The image processing and stacking}\label{proc_stack}
Our stacking technique consists in averaging the images of a large
number of galaxy clusters after masking all the unwanted sources. Since we are interested
both in the diffuse emission from intracluster stars and in the overall cluster luminosity,
including galaxies, two different masks must be utilised. In the
first case we mask all the detectable sources excluding the BCG (masks ``A''), while the foreground
stellar sources
only are masked when studying the total emission (masks ``B''). In this second case,
contamination from galaxies
not belonging to the clusters is significant. However, thanks to the large number of fields
that are stacked, their light is almost uniformly distributed in the stacked frame and
their contribution to the surface brightness can be reliably estimated sufficiently far from
the centre of the cluster.
Detection limits and surface brightness thresholds in building the ``A'' masks are expected 
to have an important influence on the estimated amount of diffuse light: we will thoroughly 
discuss this issue in Sections \ref{corrections} and \ref{systematics_mask}.

A big advantage of the stacking approach with respect to traditional imaging of individual
clusters resides in the possibility of applying very conservative
masking to the foreground (stellar) sources, without any significant loss in the measured
signal. Combined with the uniform distribution of contaminating foregrounds
in the resulting stacked image, this masking makes a careful modelling of the point 
spread function (PSF) unnecessary in order to subtract foreground stars.\\

The SDSS imaging data are available as bias subtracted, flat-field corrected frames; 
adopting the standard SDSS terminology, we will refer to these as ``corrected frames''
in the following.
We use only the three most sensitive SDSS pass-bands, $g$, $r$ and $i$.
Before images can be actually stacked, background subtraction, geometric transformation
and intensity rescaling must be applied.
For each cluster we estimate the sky background in an annulus with inner radius corresponding
to 1 Mpc, 100 kpc thick, centred on the BCG. Sources lying in that area were masked
using the segmentation image obtained by running \textsc{SExtractor} 
\citep[][]{sex}\footnote{We use \textsc{SExtractor} version 2.3 throughout the present work.}
with a Gaussian smoothing kernel (FWHM = 4.0 pixels), and a detection threshold of 0.3 times
the local rms and 5 pixels as minimum area.

We extract $1600\times1600$ pixel$^2$ ($\approx 634\times634$ arcsec$^2$) frames centred on 
the BCG, using the standard
\textsc{iraf} task \texttt{geotran}. Pixels are resampled using linear interpolation and
all images are rescaled to the same metric length at the redshift of the cluster.
Considering the redshift distribution of the sample, we rescale the images such that
$1~\mathrm{Mpc}=629.3~\mathrm{pixels}$, thus minimising the average rescaling.
In addition we either apply a random rotation before stacking, or we align the images
based on the BCG orientation. The first method is more appropriate to study the radial
surface brightness profiles, since it ensures the perfect central symmetry of the stacked
images. Aligning the images to the BCG is most suitable in order to study the
correlation between the 2-dimensional shapes of different luminous components, as will be shown in
Sec. \ref{ellipticity_sec}.

Pixel counts are then rescaled in order to remove the effects of the variation of Galactic 
extinction and cosmological surface brightness dimming $(1+z)^4$ between different clusters, 
and to homogenise the photometric calibration \citep{fukugita_etal96,hogg_etal01,smith_etal02}.
This was done according to the following equation:
\begin{equation}
c^\prime = c\times \frac{f_{20,\mathrm{ref}}}{f_{20}}\times10^{0.4\times A_\lambda}\times 
\left(\frac{1+z}{1+z_{\mathrm{ref}}}\right)^4\label{phototrans}
\end{equation}
where $c^\prime$ and $c$ are the counts in a pixel after and before intensity rescaling,
respectively; $f_{20,\mathrm{ref}}$ and $f_{20}$ are the counts corresponding to 20 mag
in the arbitrary reference calibration system and in the frame calibration system respectively;
$A_\lambda$ is the Galactic extinction as reported in the SDSS database, according to
\cite{schlegel_dust}; $z$ and $z_{\mathrm{ref}}$ are the redshift of the cluster and the 
reference redshift, that has been chosen to be close to the median redshift of the sample,
that is 0.25. This calculation ignores K-corrections as they are unknown for the ICL and
given the small redshift range
probably have negligible effects on our results. In the following we will always use the subscript
``0.25'' to refer to 
magnitudes and surface brightness in the photometric system defined by Equation \ref{phototrans}.\\

Masks ``A'' and ``B'' are built for each cluster from an analysis of the original corrected
frames, and then geometrically transformed to match the corresponding images.
First we build the ``B'' mask for the saturated sources and the stars in the field.
Relying on the SDSS photometric database, we select all objects which are flagged as saturated
and lie within 10 arcmin of the BCG. These are masked out to an extent of three times the
maximum isophotal radius in the three bands, in order to avoid including scattered light or their
bright extended haloes in the stack. All stars brighter than 20 mag in $r$-band, with
isophotal radius measured in at least two bands,
are identified and masked out to their maximum isophotal radius in the three bands.
The magnitude limit is chosen such that less than 1 per cent of objects classified
as stars by the photometric pipeline are likely to be misclassified galaxies
\citep[see][]{ivezic_stars}. This allows us to minimise the foreground signal, while
losing a negligible fraction of light from cluster galaxies. However, a small fraction (less
than 10 per cent) of bright non-saturated stars are left unmasked because they do not have
good isophotal measurements. Since these stars are randomly located in the frames, no
systematic effects on our measurements are expected, although this failure increases the noise in 
the foreground signal.

To obtain the ``A'' mask we run \textsc{SExtractor} on the frames in the three bands,
using a Gaussian smoothing kernel (FWHM = 4.0 pixels), a minimum detection area of 10 pixels
and detection thresholds corresponding to $\mu_{r,0.25} = 24.5$, $\mu_{g,0.25} = 25.0$ 
and $\mu_{i,0.25} = 24.0$ mag~arcsec$^{-2}$. We blank the segment corresponding to the BCG
in the segmentation images and \textsc{or}-combine them with the ``B'' mask previously
generated to get the final ``A'' mask. Note that both the ``A'' and the ``B'' masks are
the same for all the three bands, thus allowing a consistent measurement of the colours.\\

The stacking of the images is performed using the standard \textsc{iraf} task \texttt{imcombine}.
The images in each (sub)sample are combined with a simple average of the pixel counts, excluding 
the masked pixels.
We do not apply any kind of statistical rejection to the pixel count distributions, since
we cannot assume that the light follows the same distribution in all
clusters on a few pixel scale. This is obviously not the case when considering the total 
light which is dominated by the galaxies, but even for the diffuse component significant 
substructure can be present as well \citep[see e.g.][]{gregg_west98,trentham_mobasher98,calcaneo+00}.

\section{The photometric analysis}\label{photo_analysis}
\begin{figure*}
\centerline{
\includegraphics[width=8.5truecm]{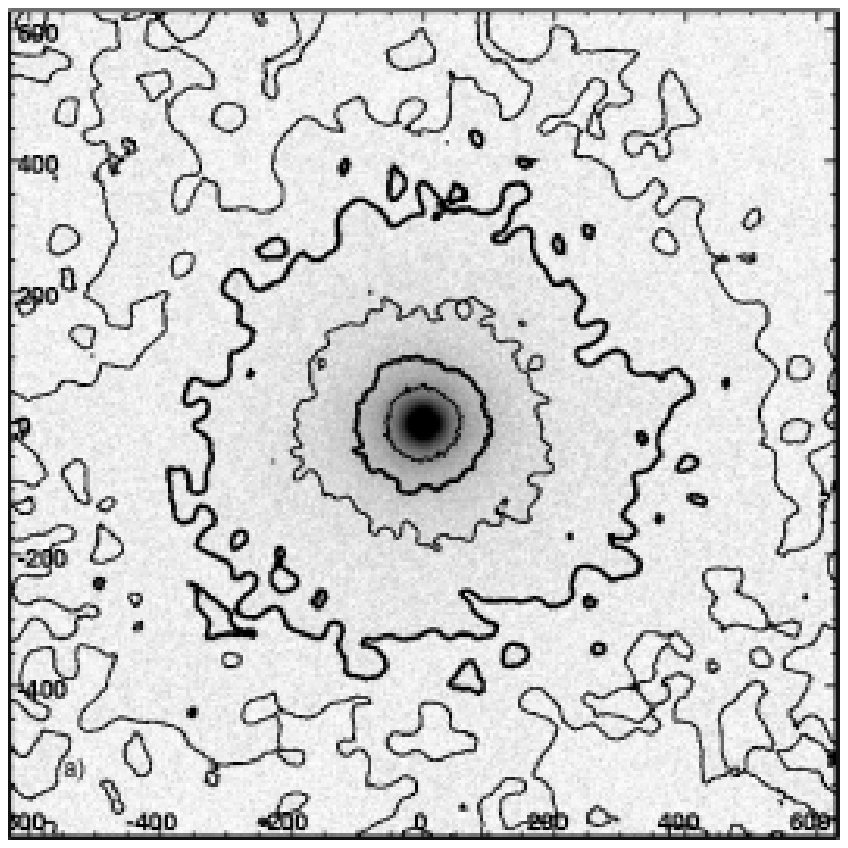}
\includegraphics[width=8.5truecm]{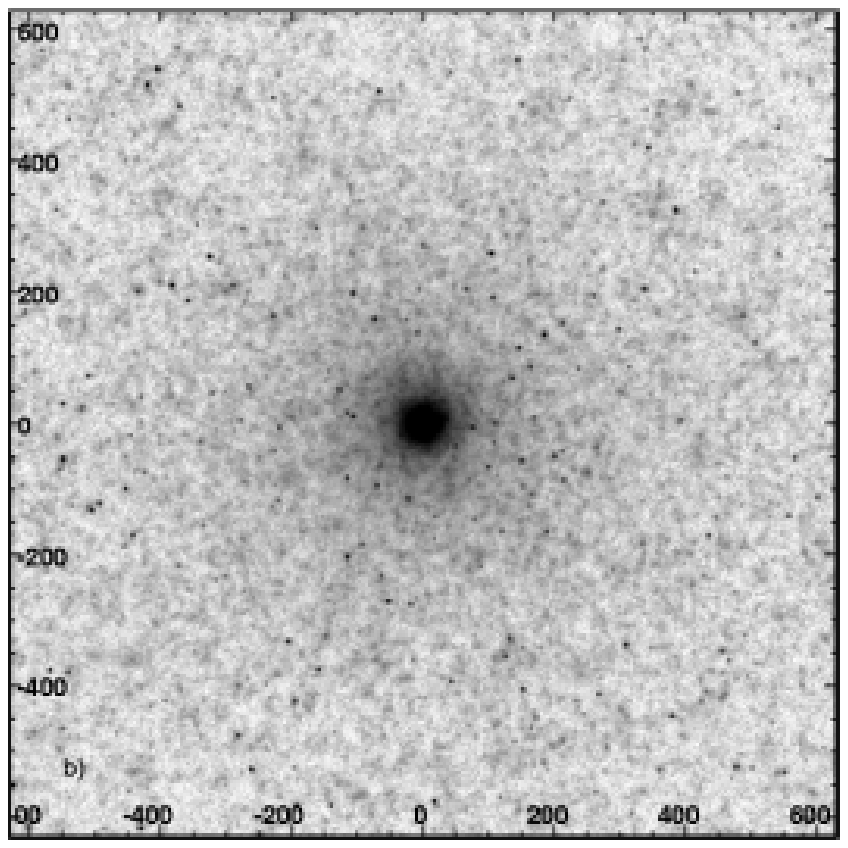}
}\caption{The $r+i$ composite images resulting from the stacking of the main sample: the
diffuse component plus BCG is in panel a), the total light in panel b). The same logarithmic grey
scale is adopted in both images. Side-scale tickmarks display the distance in kpc from the centre.
Isophotal contours corresponding to $\mu_{(r+i),0.25}$ of 26, 27, 28, 29 and 30 mag~arcsec$^{-2}$ for
the diffuse component are overplotted on panel a). Smoothing kernels of 3, 7, 11, 17, and 21
pixels respectively are used. Corresponding SB values in $r$-band are $\sim 0.3$ mag brighter.
Note the point-like sources in panel b), that are
the stars left unmasked due to the failure in the measurements of their isophotal radii, as 
mentioned in Section \ref{proc_stack}.
}\label{stack_pictures} 
\end{figure*}

In Fig. \ref{stack_pictures} we present the central $\sim 600$ kpc of the stacked images 
in the total sample, for the
diffuse component plus BCG (``A'' masks, left panel) and for the total emission (``B'' masks, right 
panel). In order to increase the $S/N$, the $r$ and $i$ bands
have been combined in these plates using a weighted average of the intensities, 
where the weights are 
given by the inverse square of the rms of the intensity in each stacked pixel.
The combined intensities are 
translated into AB mag for a combined $r+i$ pass-band, whose effective response function is given
by the sum of the two filter responses.
Scales are marked in kpc. In panel a) we superpose the isophotal contours
corresponding to 26, 27, 28, 29 and 30 mag~arcsec$^{-2}$ in the $r+i_{0.25}$ band, obtained 
smoothing the original image with kernels of increasing size, as described in the caption.
Corresponding SB values in $r$-band are $\sim 0.3$ mag brighter.

In both panels the central region ($R\lesssim 100$ kpc) is dominated by the BCG, which
has not been covered by either the ``A'' or ``B'' masks. Given the circular symmetry of the stacked images 
and the need to integrate over large areas to increase the $S/N$, we perform the photometric
analysis in circular apertures.
The stacked image is first divided into a number of circular annuli, centred on the BCG and
logarithmically spaced. Within each annulus, the average surface brightness is computed
simply by summing the intensity in the pixels and dividing by the total number of pixels.
The logarithmic spacing ensures that a larger area is summed at large radii, where the
signal is lower. In order to evaluate the statistical uncertainty on the computed SB, we further
divide each annulus into a number of sectors with aperture angle $\theta\simeq\frac{\Delta R}{R}$,
where $\Delta R$ is the thickness of the annulus and $R$ its average distance from the centre.
The rms of the SB among the sectors is thus representative of the SB fluctuations on the typical
spatial scale covered by the annulus; the statistical error on the average SB is then just given 
by the rms divided by $\sqrt{N-1}$, $N$ being the number of sectors in the annulus.

Background estimation is a critical issue when attempting to measure very low surface brightnesses,
which are just a few sigma above the noise. In particular, when integrated fluxes over
large areas are estimated, the uncertainty on the background level dominates the measurement error.

Due to the limited spatial coverage provided by the individual SDSS frames, at $R\gtrsim 1$ Mpc and
beyond, the fraction of stacked images which contribute to the intensity of each pixel drops
below 50 per cent.
Given the average $R_{200}$ of 1.1 Mpc for the clusters in the main sample,\footnote{This 
is consistent with a mean mass of 7--8$\times 10^{13} M_\odot$.} we cannot directly estimate the 
background level around the stacked cluster images.
Since the mean galaxy profile of clusters is known to follow the analytical profile proposed by
\cite{NFW} \citep[NFW, see, for example,][]{NFW_CNOC}, we estimate the background SB level
in the stacked images by fitting a projected NFW profile \citep[see ][]{bartelmann_96} plus a 
constant to the SB profiles extracted before. The fitting is performed between 100 and 900 kpc,
the innermost 100 kpc being excluded because of the predominance of the de Vaucouleurs profile
\citep{dev_profile} of the BCG in these central regions. The best fitting value of the constant is used as the 
residual background level, which may be positive or negative, since a sky background has already
been subtracted from each individual cluster image based on the mean sky surface brightness
1 Mpc away from the BCG. The corresponding 
uncertainty is given by the square root of its variance as determined from a set of Monte Carlo 
realizations of the measured profile, where the intensity in each point is randomly varied according 
to the associated error. 
In Section \ref{discussion} we will discuss in further detail the validity of this method.

\subsection{Definition of the ICL and corrections for mask incompleteness}\label{corrections}
A measurement of the intracluster light, conventionally defined as the luminous emission from 
stars which are unbound to any individual galaxy, would require a fully dynamical characterisation
of such stars. This is completely beyond our observational capabilities. Based on purely 
photometric properties, the most sensible and robust definition of the ICL
is the emission coming from outside the optical boundaries of individual galaxies.
Given the properties of our dataset, we define the optical boundary as the
$\mu_{r,0.25}=25.0$ mag~arcsec$^{-2}$ isophotal contour. This does not encompass the low
surface brightness emission from the outermost parts of galaxies, so a fraction
of what we define as ICL will actually come from stars bound to galaxies, as we will argue more
quantitatively in Sec. \ref{ICLgal_sec}.\\

In addition to this contribution which is inherent to our definition of the ICL,
contamination arises also from incomplete masking (i.e. from masks failing to cover the entire
optical extent of a galaxy) and from undetected galaxies. 
We have tested the efficiency of our masking algorithm on a mock dataset of $\sim 650$ 
simulated clusters. In order to reproduce the observational properties of the real
sample, we use the same redshift, Galactic extinction, background noise properties, 
point spread function (PSF) and photometric calibration parameters as in the real cluster sample.

For each simulated cluster, we generate 1,000 random galaxies whose photometric properties are 
assigned as follows. First we draw their absolute $r$-band luminosity from a Schechter luminosity 
function (LF), using the fitting parameters of \cite{mobasher_lf} ($\alpha=-1.18$, 
$M_R^*=-21.79+5 \log h_{65}$)\footnote{This corresponds to $M_r^*=-21.37+5 \log h_{70}$
using the photometric conversion provided by the \cite{BC03} code for a 13 Gyr old, solar 
metallicity SSP.}, 
who studied the LF of the Coma cluster down to $M_R=-16+5 \log h_{65}$. Using the data in 
\cite{blanton03_sersic},
we have computed the two-dimensional conditional probability function $P(n,\mu_e|L)$
of the S\'ersic index $n$ and $\mu_e$ for a given luminosity $L$. According to this,
we assign random $n$ and $\mu_e$ to each galaxy. A projected axial ratio is randomly drawn
from a Gaussian distribution centred at 0.7 and with $\sigma=0.25$ \citep[we assume
that clusters are dominated by early-type galaxies and use the results in][]{lambas+92},
the orientation is random.

The model SB distribution in the $r_{0.25}$-band is then computed {\it out to the optical radius}
(as defined above).
Size and intensity are rescaled according to the redshift\footnote{We adopt a uniform 
K-correction within each cluster, namely the one for a spectral energy distribution of a
simple stellar population formed 10 Gyrs ago with solar metal enrichment \citep{BC03}.},
Galactic extinction and
photometric calibration of the corresponding observed cluster. The model distribution is
sampled on the pixel array. The resulting image array is then convolved with the PSF and
the sky background is added. Finally we add Poisson noise, using the actual electron-to-ADU
conversion factor of each frame. We use the average colours of the total light in the stacked 
images to derive from the $r_{0.25}$-band model image the corresponding mock frames in the 
$g$ and $i$ bands.

The mock images are processed with the same codes and algorithms used for the real ones, and
stacked. We find that roughly 15 per cent of the light in the $r$-band within the optical radius 
of the simulated galaxies fails to be blocked by our masking algorithm, because of partial or complete
non-detections. Very similar amounts are missed in $g$ and $i$.
These results clearly are expected to depend on the shape of the luminosity function of the cluster
galaxies, and particularly on the slope at the faint end. In fact a steeper faint-end and/or
a fainter $M^*$ can produce significantly higher contaminating fractions, and {\it vice versa}.
We will discuss the systematic uncertainty deriving from the choice of the LF in further detail
in Sec. \ref{systematics_mask}. Throughout the rest of the paper we will adopt the corrections based on 
the \cite{mobasher_lf} LF as our fiducial corrections.

Assuming that the fraction $f$ of unblocked galaxy luminosity is roughly independent of
clustercentric distance, we can compute the corrected surface brightness of the ICL:
\begin{equation}
\Sigma_{\mathrm{ICL}}=\frac{1}{1-f}\times\Sigma_{\mathrm{diffuse}}-\frac{f}{1-f}\times\Sigma_{\mathrm{total}}
\label{ICLcorr}\end{equation}
where $\Sigma_{\mathrm{diffuse}}$ and $\Sigma_{\mathrm{total}}$ are the surface brightness
as derived from the stacked images masked with the ``A'' and ``B'' masks respectively.

\section{Results}\label{results}
\subsection{Surface brightness profiles}
We present the results of our photometric analysis for the main sample in Fig. \ref{profiles}.
In the first three panels (a) to (c) we show surface brightness profiles
for various cluster components (lower plots) as well as the local ratio of $ICL+BCG$ to
the total cluster light 
(upper plots) for the $g$-, $r$- and $i$-band. The radial coordinate is scaled to $R^{1/4}$.
In the SB profile plots, red triangles with error bars represent the average \textit{total} SB,
including all cluster components, whereas black open circles are the SB of the \textit{diffuse}
component ($ICL+BCG$). The SB of the \textit{ICL+BCG} component after correcting 
for the contamination due to mask incompleteness is shown by black filled circles with error 
bars.
For the \textit{total} SB error bars represent the maximum range of variation when allowing
for 1-$\sigma$ uncertainties in the background level and in the local estimate of the SB.
For the \textit{ICL+BCG} component the analogous maximum range of variation is computed
taking errors on both \textit{total} and \textit{diffuse} light measurements into account, after
combining them with the standard error propagation formulae applied to Equation \ref{ICLcorr}.
The two horizontal lines display the SB corresponding to 1-$\sigma$ background uncertainty,
for the \textit{total} light (red dashed line), and for the \textit{diffuse} light (black solid line).

The profiles can be reliably traced out to $\sim$700 kpc at the level
of $\mu_{r,0.25}\sim32$ for the ICL ($\mu_{r,0.25}\sim31$ for the uncorrected 
\textit{diffuse} component) and $\mu_{r,0.25}\sim29$ mag arcsec$^{-2}$
for total light (5-$\sigma$ detections are obtained at 500 kpc in all bands).

As apparent from the straight-line behaviour in all three bands, in the inner 40--50 kpc the 
SB profile of the \textit{ICL+BCG} component closely follows a de Vaucouleurs' law. We fit this 
law to the inner points of the profiles using an iterative procedure
based on standard least squares.
We start fitting between 10 and 20 kpc to ensure that we exclude the innermost region, where the 
seeing (FWHM$\sim 5$ kpc at the average redshift of the clusters) is likely to significantly affect 
the measured profile. Outer points are added iteratively until the error on the slope
is increased by more than 10 per cent with respect to the previous step. The last two points are
then discarded and the fitting parameters are recomputed.\\
The best fitting de Vaucouleurs' law to 
the inner \textit{ICL+BCG} profile is plotted as a dotted line in the lower section of panels a) to 
c) of Fig. \ref{profiles}. The effective radius $R_e$ is marked with a vertical dashed line and
fitting parameters are reported nearby. Typical errors are $\sim 10$ per cent on $R_e$ and
$\sim0.1$ mag arcsec$^{-2}$ on $\mu_e$. We note that, while the effective radii are consistent
in $r$ and $i$, in the $g$-band $R_e$ is significantly larger by some 5 per cent.
This is consistent with a significant colour gradient in the expected sense within the BCG itself.

Beyond 50 kpc the profiles clearly flatten, both in the \textit{ICL} component and in the 
\textit{total} light. Still, there is evidence for the \textit{ICL} component being more 
centrally concentrated. The $R^{1/4}$ slope of the \textit{ICL} component stays roughly
constant between 150 and 500-600 kpc, with an equivalent $R_e$ of 250--300 kpc in all
three bands. The $total-diffuse$ light is also quite well fit by an $R^{1/4}$ law, but here the
equivalent $R_{\mathrm e}$ is $\sim2$ Mpc in all three bands.

The higher concentration of the \textit{ICL} component with respect to the \textit{total} light is
confirmed by the upper sections of the plots in Fig. \ref{profiles}, where we plot the local 
ratio of
the \textit{ICL+BCG} to the \textit{total} light using black filled circles with error bars. 
Error bars are computed by combining the errors on the fluxes
derived as described above. Empty circles represent the fraction of uncorrected \textit{diffuse}
light, that is the upper limit to the ICL fraction when one allows for different
choices of the LF.\\
Excluding the central $\sim50$ kpc, where the BCG
dominates, a smooth monotonic decrease of the \textit{ICL} fraction is observed all the way out
to the limit of significant detections, from $\sim50$ per cent at 50 kpc down to $\lesssim$5
per cent at 600--700 kpc for our preferred faint galaxy correction.
By comparing the two fractions, we see that the \textit{ICL} is the main component of the 
\textit{diffuse} light out to $\sim$300 kpc. Outside this radius we must expect that 
the \textit{diffuse} emission reflects more and more the properties of faint cluster galaxies 
rather than those of the ICL.\\

\subsection{Colours of the ICL}
\begin{figure*}
\centerline{
\includegraphics[width=8.5truecm]{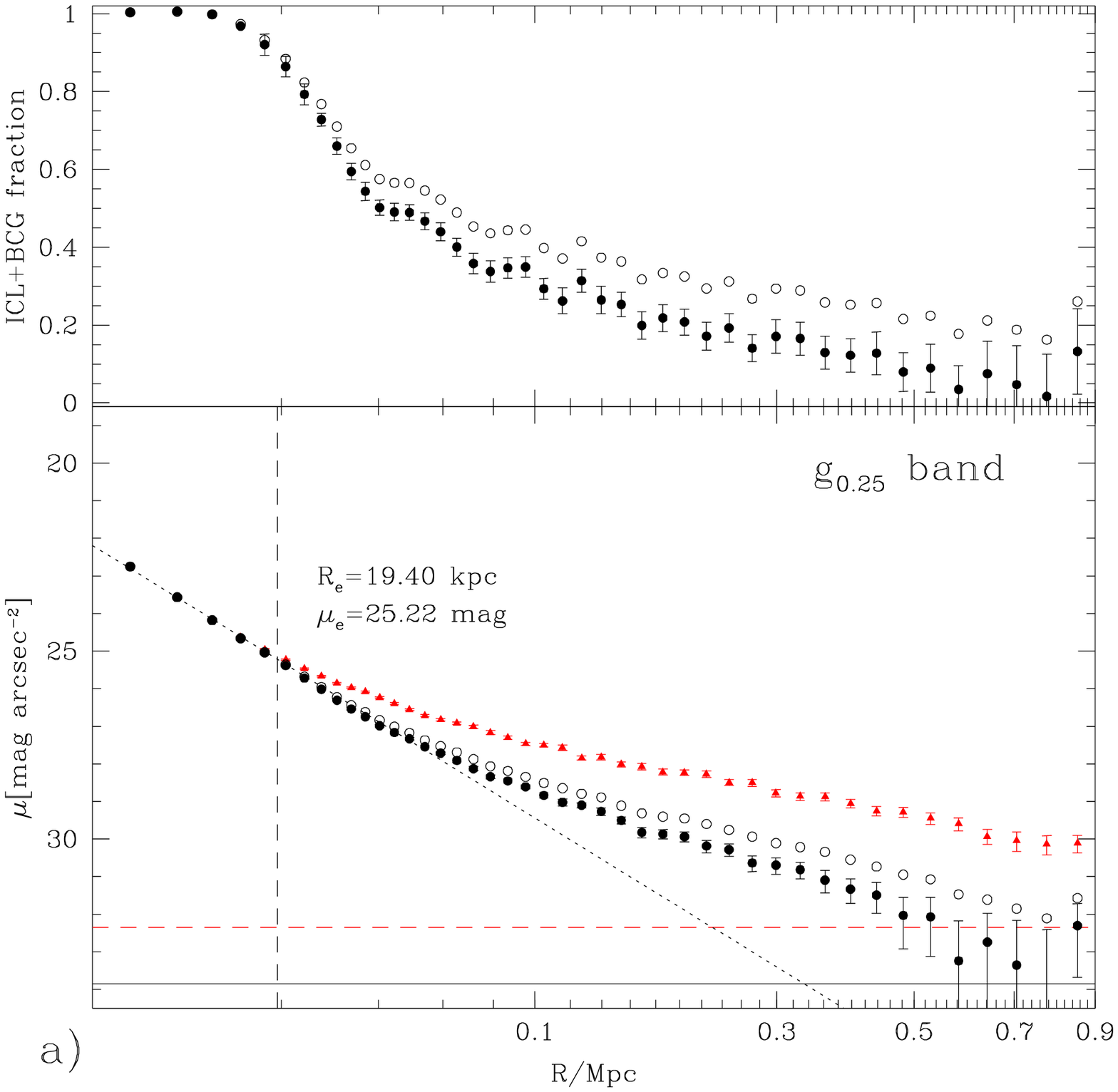}
\includegraphics[width=8.5truecm]{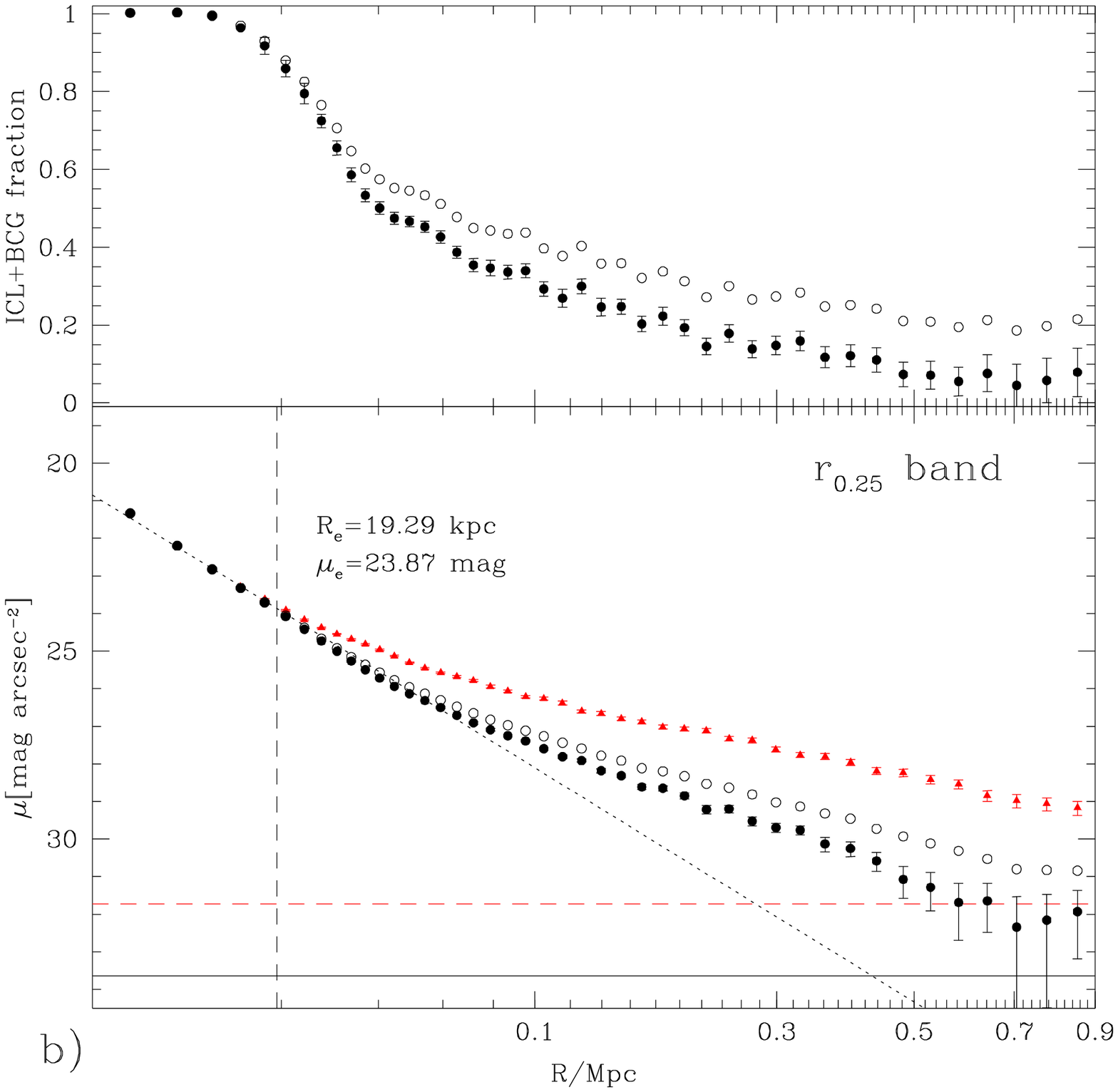}
}
\centerline{
\includegraphics[width=8.5truecm]{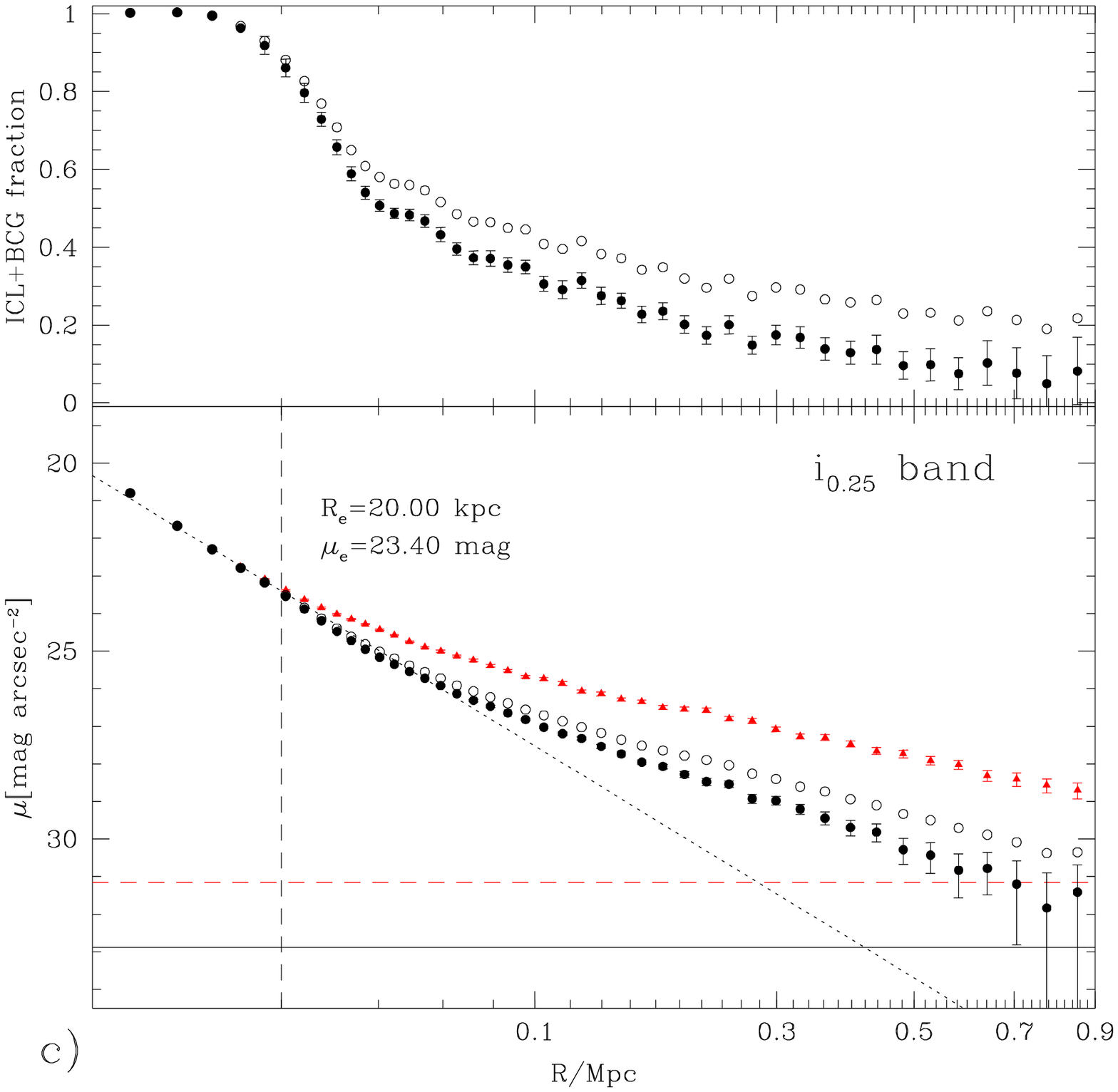}
\includegraphics[width=8.5truecm]{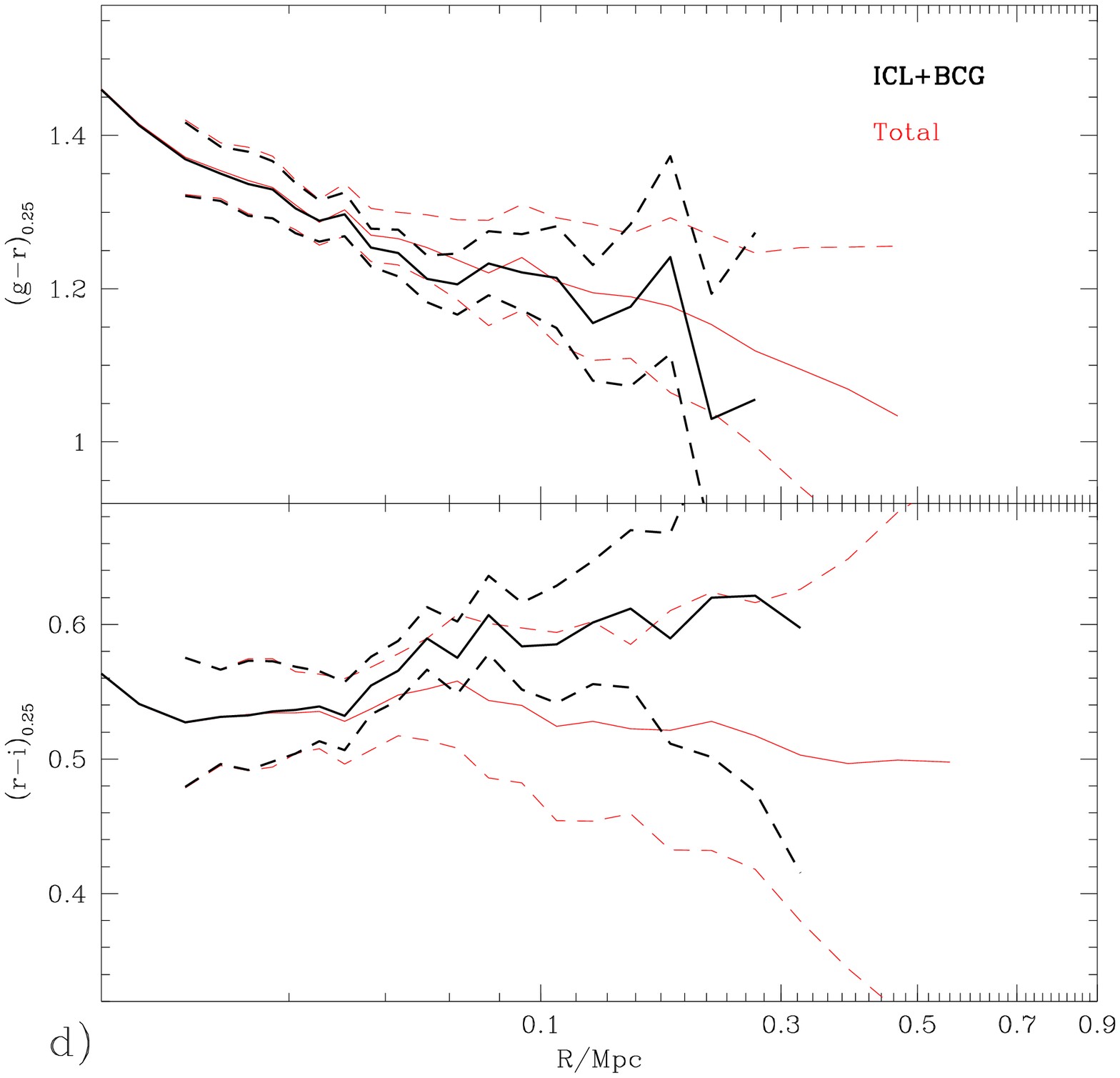}
}
\caption{Surface photometry of the main sample.
Panels a) to c) show the surface brightness profiles and the local ratio of \textit{ICL+BCG}
and uncorrected \textit{diffuse} light
to the {\it total} cluster light in the $g$, $r$, $i$ bands respectively; in panel d) the $g-r$ 
and $r-i$ colour profiles are plotted. The $R$ axis is linear in $R^{1/4}$.
Panels a)--c), lower section: the SB is expressed in mag arcsec$^{-2}$ in the $z=0.25$ observer
frame. Red triangles with error bars represent the total cluster light, open circles
the diffuse light (including the BCG) as directly measured from the stacked images. Filled circles
with error-bars display the SB of the {\it ICL+BCG}, corrected for masking incompleteness
adopting the LF given by Mobasher et al. (2003). Horizontal red dashed and black solid lines display 
the SB corresponding to the
1-$\sigma$ uncertainties on the background determination for the total light and for the 
{\it ICL+BCG} respectively. The dotted lines represent the best de Vaucouleurs fits to the inner
regions (see text for the details): the effective radii of the best fitting models are indicated 
with vertical dashed lines and the corresponding parameters are reported nearby.
Panels a)--c), upper section: the local ratio of {\it ICL+BCG} (filled dots with error-bars)
and uncorrected \textit{diffuse} light (open circles) to {\it total} cluster light. 
Panel d):  $g-r$ (upper section) and $r-i$ (lower section) colour profiles. Red thin lines are used for the
{\it total} light, black thick lines for the {\it ICL+BCG} component, corrected according to the 
Mobasher et al. (2003) LF. Dashed lines represent the 1-$\sigma$ confidence intervals.}\label{profiles}
\end{figure*}

In panel \ref{profiles}(d) we show the $g-r$ and $r-i$ colour profiles for the ICL+BGC (black thick solid lines) and for 
the total light (red thin solid lines). Dashed lines show the 1-$\sigma$ confidence interval
derived from the errors on the fluxes in the different bands. Only data-points for which
the confidence interval is smaller than 0.5 mag are connected, in order to avoid confusion
arising from noisy measurements. We observe a striking consistency between the colours of the
{\it total} emission and those of the {\it ICL+BCG} in both $g-r$ and $r-i$ out to 300 kpc, where 
the light from bound stars in galaxies dominates the {\it total} emission. There is marginal 
evidence for slightly redder $r-i$ (+0.03 mag) in the ICL at $R>100$ kpc, but its significance 
is low, less than 1-$\sigma$.

In agreement with the above determinations of $R_e$, within 80--100 kpc we find a clear gradient in 
$g-r$, from 1.4--1.5 in the centre to $\sim1.20$ at 80 kpc, whereas the $r-i$ profile is consistent
with being flat over this radius range, within the errors, with an average value $\sim0.60$. 
Outside 80 kpc the $g-r$ profile flattens too, both for \textit{ICL+BCG} and  for the 
\textit{total} light.\\
\\  

\subsection{Isophotal ellipticity}\label{ellipticity_sec}
We now investigate possible relationships between the shape of the BCG and those of the ICL and 
of the cluster galaxy distributions. We study the isophotal shapes of the
ICL and of the galaxy light for two subsamples of clusters in which the BCGs exhibit a clear
elongation. The stacking in this case has been performed after aligning the images along the major
axis of the BCG. The selection is from the main sample clusters after requiring 
$m_{r,0.25}$(BCG)$<17.90$
to ensure a more reliable estimate of the parameters of the best fitting 2-dimensional
de Vaucouleurs model, as provided by the SDSS \texttt{photo} pipeline \citep{lupton_etal01}. 
We choose an axial ratio limit of
${b/a}_{\mathrm{BCG}}<0.85$, for the first subsample, and a more restrictive limit of 
${b/a}_{\mathrm{BCG}}<0.70$, 
for the second one. The images are aligned according to the position angle provided by
\texttt{photo}. A few clusters, for which the fitting algorithm clearly
failed to provide a sensible description of the shape of the BCG, have been rejected after 
visual inspection. The two final samples contain 355 and 112 clusters.

The analysis of the resulting \textit{diffuse} light image is performed by fitting elliptical 
isophotes. Such isophote fitting cannot be used for the galaxy light,
because of the shot noise arising from galaxy discreteness.
We therefore characterise the shape of the galaxy distribution by means of 
moments of the $total-diffuse$ light image.
We use $r+i$ band composite images derived
from the weighted average of the two single bands (cf. Fig. \ref{stack_pictures}), to
enhance the $S/N$.

In the central 50 kpc of the \textit{diffuse} light image
the fitting is done using the standard \textsc{IRAF} task \texttt{ellipse}. Outside this radius
the $S/N$ is too small to make the fitting 
method implemented in \texttt{ellipse} applicable. In fact, this method \citep[see][]{jedre87} 
consists in minimising the variance of the intensity along 1-pixel wide elliptical paths by 
varying the geometrical parameters of the ellipse. In our fitting code the elliptical paths are
replaced with elliptical annuli, which are several pixel wide, and the intensity variance
is computed not on single pixel basis, but using approximately square (side$\sim$annulus width),
non-overlapping contiguous regions along the annulus. For each given semi-major axis $a$
the width of the elliptical annulus $\Delta a$ is fixed as $a\times 0.2$. The ellipse is
aligned with the BCG. We compute the axial ratio $b/a$ that minimises the intensity variance
using the standard golden section search algorithm, as implemented in \cite{NR}.
\begin{figure}
\includegraphics[width=8.5truecm]{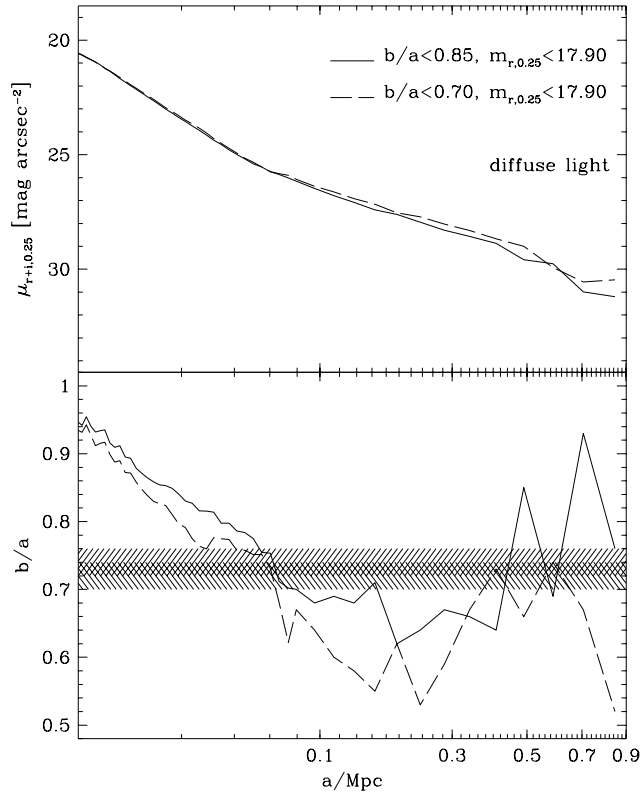}
\caption{Elliptical isophote analysis of the diffuse light of two subsamples of clusters, 
whose images have
been aligned along the position angle of the best-fitting de Vaucouleurs model for the
BCG. Only clusters with luminous BCG ($m_{r,0.25}({\mathrm{BCG}})<17.90$) have been
stacked. As a function of the semi-major axis $a$ we show the
isophotal SB in the $r+i$ composite band (upper panel) and the axial ratio $b/a$ (lower panel).
Solid lines are used for the sample with ${b/a}_{\mathrm{BCG}}<0.85$, dashed lines 
for ${b/a}_{\mathrm{BCG}}<0.70$. The shaded regions indicate the mean ellipticity value (plus
uncertainty)
of the galaxy distribution, with the two different slants representing the two subsamples.}\label{ellipticity}
\end{figure}\\

The results of this analysis are shown in Fig. \ref{ellipticity}, where the two panels
display the average SB of the isophotes (upper panel) and the axial ratio $b/a$ 
of the isophotal ellipses (lower panel). Solid lines are for the sample with 
${b/a}_{\mathrm{BCG}}<0.85$, while dashed lines are for ${b/a}_{\mathrm{BCG}}<0.70$. 
Typical errors for $b/a$ range from 0.05 to $\gtrsim 0.10$ going from the centre to
700 kpc.
The two samples display very similar behaviour, except for a lower $b/a$
in the sample with flatter BCGs, as expected. We note a progressive 
flattening of the isophotes from the centre out to $a\sim150$ kpc. While the smearing effect 
of the PSF (FWHM $\sim5$ kpc) is certainly biasing the measurements in the central 10--20 kpc,
at larger distances this flattening must be considered real.
At $a\gtrsim$500 kpc, large $b/a$ uncertainties ($\gtrsim 0.1$) do not allow us to establish
whether the apparent increasing trend is real or not. As noted above, however, the \textit{diffuse} 
light is expected to be dominated by unmasked galaxy light at these distances, and therefore does 
not provide reliable information on the distribution of the ICL.\\

The flattening of the cluster galaxy distribution is derived from  first order
moments of the $(total-diffuse)$ light image, $S_x=\sum_i |x_i| F_i$ and $S_y=\sum_i |y_i| F_i$,
where the $x$ and $y$ coordinates are aligned with and perpendicular to the BCG orientation 
respectively, $F_i$ is the fraction of light in pixel $i$, and the sums are within an ellipse having
700 kpc semi-major axis and axial ratio $b/a$. The $b/a$ that best fits the flattening of the galaxy
light distribution is obtained by requiring
$\frac{S_y}{S_x}=b/a$. For our two subsamples we obtain $b/a=0.74\pm0.02$ and $b/a=0.72\pm0.02$.
These results do not change appreciably for variations of the semi-major axis $a$ in the range
700 to 350 kpc. The galaxy distribution is thus significantly rounder than that of the ICL
between 100 and 400 kpc. In fact, by using the \textit{diffuse} light as proxy for the ICL,
we are probably underestimating the real effect. Unless the faint galaxy 
population that contaminates the \textit{diffuse} light has significantly higher degree of alignment
with the BCG than the bright one, the ellipticity we measure in the \textit{diffuse} image should
be smaller than in the real ICL.

It is interesting to note that the semi-major axis at which the maximum flattening of the isophote
is first reached, $a\sim 100$ kpc, lies somewhat outside the radius where the outer shallower SB
profile takes over from the inner de Vaucouleurs and the 
galaxy component starts dominating the total light.

\subsection{ICL--galaxy connection}\label{ICLgal_sec}
We further investigate the connection between the galaxy distribution and the ICL by 
studying the photometric properties of the ICL in regions at different projected distances from 
bright galaxies.
First we select all the galaxies more luminous than $M_r^*+2.0$ mag in the SDSS
database; this corresponds to $m_{r,0.25}<21.23$ assuming the \cite{mobasher_lf} LF.
All pixels contributing to the {\it diffuse} light image (i.e. not masked according to
type ``A'' masks) are partitioned into 4 different classes 
according to their distance $l$ from the nearest bright
galaxy, namely $l<15$ kpc, $15\leq l<25$, $25\leq l<40$, and $l>40$ kpc. We further distinguish
between 4 different ranges in clustercentric distance: from 100 to 200 kpc, from 200 to 400,
from 400 to 750 and from 800 to 900 kpc, this last one being roughly representative of the 
background.
Fluxes in each class of $l$ and clustercentric distance are stacked separately.
As the best estimate of the background level we adopt in this case the average surface brightness 
of pixels in the 800--900 kpc annulus with $l>40$ kpc.

\begin{figure}
\centerline{
\includegraphics[width=8.5truecm]{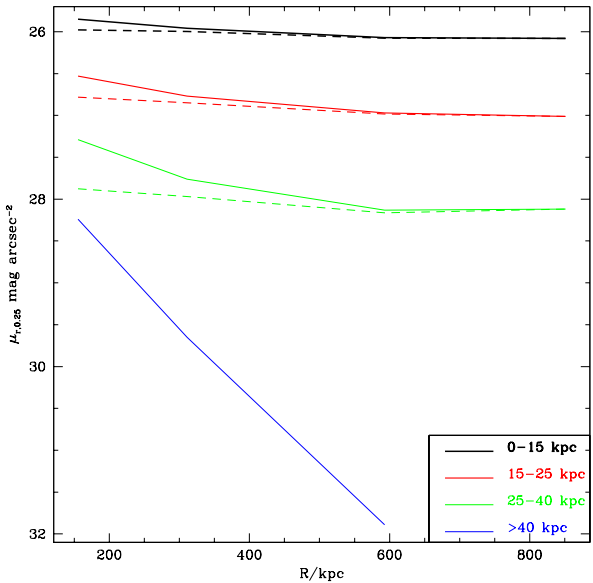}
}
\caption{SB profiles of regions at different distances from bright galaxies.
Colours correspond to four distance bins, as indicated in the legend. Dashed lines represent
the SB excess with respect to regions at distance $l>40$ kpc.}\label{ICLgal_fig}
\end{figure}

The results of this stacking are reported in Table \ref{ICLgal}. The clustercentric distance $R$
(in kpc) is given in column (1); column (2) reports the projected distance $l$ to the nearest
bright galaxy (in kpc);
the average SB of these pixels is given in column (3), while column (4) displays the fraction of ICL
in each annulus contributed by pixels in the specific $l$--range.
For the first three classes of $l$ we compute the difference between the mean flux 
actually measured and the flux that would be measured if those pixels had the same SB as the ones 
at $l>40$.
The ratio of this flux excess to the {\it total} ICL flux in the annulus of clustercentric 
distance $R$ is reported in column (5). Finally, the fraction of pixels with projected distance 
$l$ to the nearest bright galaxy is given in column (6).
The fractions in the last three columns are given in per cent.
\begin{table}
\begin{minipage}{8.5truecm}
\caption{ICL--galaxy connection: SB and relative flux and flux excess of the diffuse light
at different distance from bright galaxies.}\label{ICLgal}
\begin{tabular}{crrrrr}
\hline
$R$&$l$ & $\mu_{r,0.25}$ & $F$ & $F_{\mathrm{excess}}$ & Area\\
kpc&kpc & mag arcsec$^{-2}$ & \multicolumn{3}{c}{per cent}\\
(1) & (2) & (3) & (4) & (5) & (6)\\
\hline
100--200&&&&\\
&0--15  & 25.85 &  5.7 &  5.0 & 0.8\\
&15--25 & 26.53 & 13.4 & 10.6 & 3.7\\
&25--40 & 27.29 & 16.7 &  9.8 & 9.3\\
&$>40$  & 28.24 & 64.2 & --   &86.2\\ 
\hline
200--400&&&&\\
&0--15  & 25.96 & 10.0 &  9.7 & 0.6\\
&15--25 & 26.77 & 20.2 & 18.8 & 2.6\\
&25--40 & 27.76 & 20.6 & 17.0 & 6.6\\
&$>40$  & 29.65 & 49.2 & --   &90.2\\ 
\hline
400--750&&&&\\
&0--15  & 26.07 & 19.4 & 19.3 & 0.5\\
&15--25 & 26.97 & 33.4 & 33.1 & 1.9\\
&25--40 & 28.13 & 29.5 & 28.6 & 4.9\\
&$>40$  & 31.89 & 17.7 & --   &92.7\\ 
\hline
800--900&&&&\\
&0--15  & 26.08 & -- & -- &\\
&15--25 & 27.01 & -- & -- &\\
&25--40 & 28.13 & -- & -- &\\
&$>40$  & -- & -- & -- &\\ 
\hline
\end{tabular}
\end{minipage}
\end{table}\\
In Fig. \ref{ICLgal_fig} we plot, with solid lines, the SB of the {\it diffuse} light in the four
classes of $l$ as a function of the clustercentric distance. For the three nearest $l$ bins, the 
dashed lines represent the SB of the excess with respect to the SB at $l>40$.
We note immediately that the SB around luminous galaxies depends very weakly on
the clustercentric distance. This
is even more evident if one considers the surface brightness excess plotted with dashed lines in
Fig. \ref{ICLgal_fig}. This light can therefore be seen as representing the stars in the unmasked
outer regions of individual galaxies.

The SB excess around galaxies is about a quarter of the total diffuse light in the 100--200 kpc
annulus. For 200--400 kpc it is almost half of the {\it diffuse} light. At 400--750 kpc it accounts
for more than three quarters of the ICL. However, a comparison
with the total flux including galaxies, shows that the excess flux within 25 kpc 
represents a fraction between 2 and 7 per 
cent of the galaxy emission, and thus is well accounted for by the corrections for
masking incompleteness described in Sec. \ref{corrections}.
The contribution from SB excesses in pixel with 25 kpc $< l <$ and 40 kpc cannot be accounted for
by masking incompleteness, both because it is inconsistent with the estimated corrections and
because of the relatively large distance from the galaxies.
We therefore conclude that our measurements are consistent with the ICL
being dominated by a diffuse emission concentrated in the inner $\sim$300--400 kpc plus a fraction
of light clustered around bright galaxies, that dominates in the outer parts.

\subsection{Dependence on cluster properties: SB profiles}\label{prop_dep}
In this section and in the following one we analyse the dependence of the distribution and 
integrated amount of the ICL on global cluster properties. The description of the subsamples is 
given in Sec. \ref{subsamples}. In order to compensate for the decrease of $S/N$
due to the smaller number of clusters stacked in each subsample, we utilise the $r+i$ composite
images, obtained as the weighted average of the final stacked images in the two bands;
the weights are given for each pixel by the inverse variance of the intensity of the
corresponding pixels in the images to be stacked.

\begin{figure*}
\centerline{
\includegraphics[width=8.5truecm]{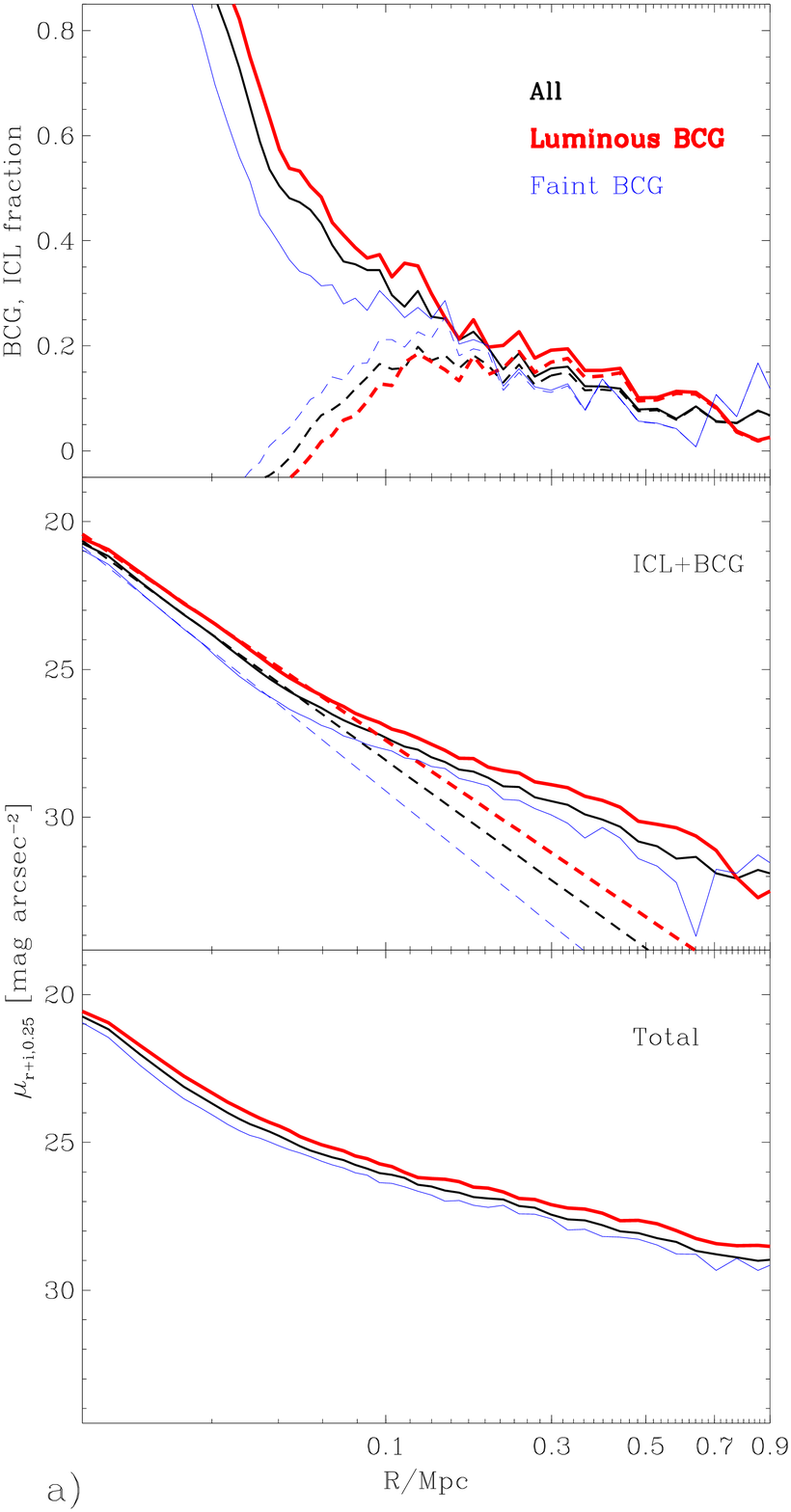}
\includegraphics[width=8.5truecm]{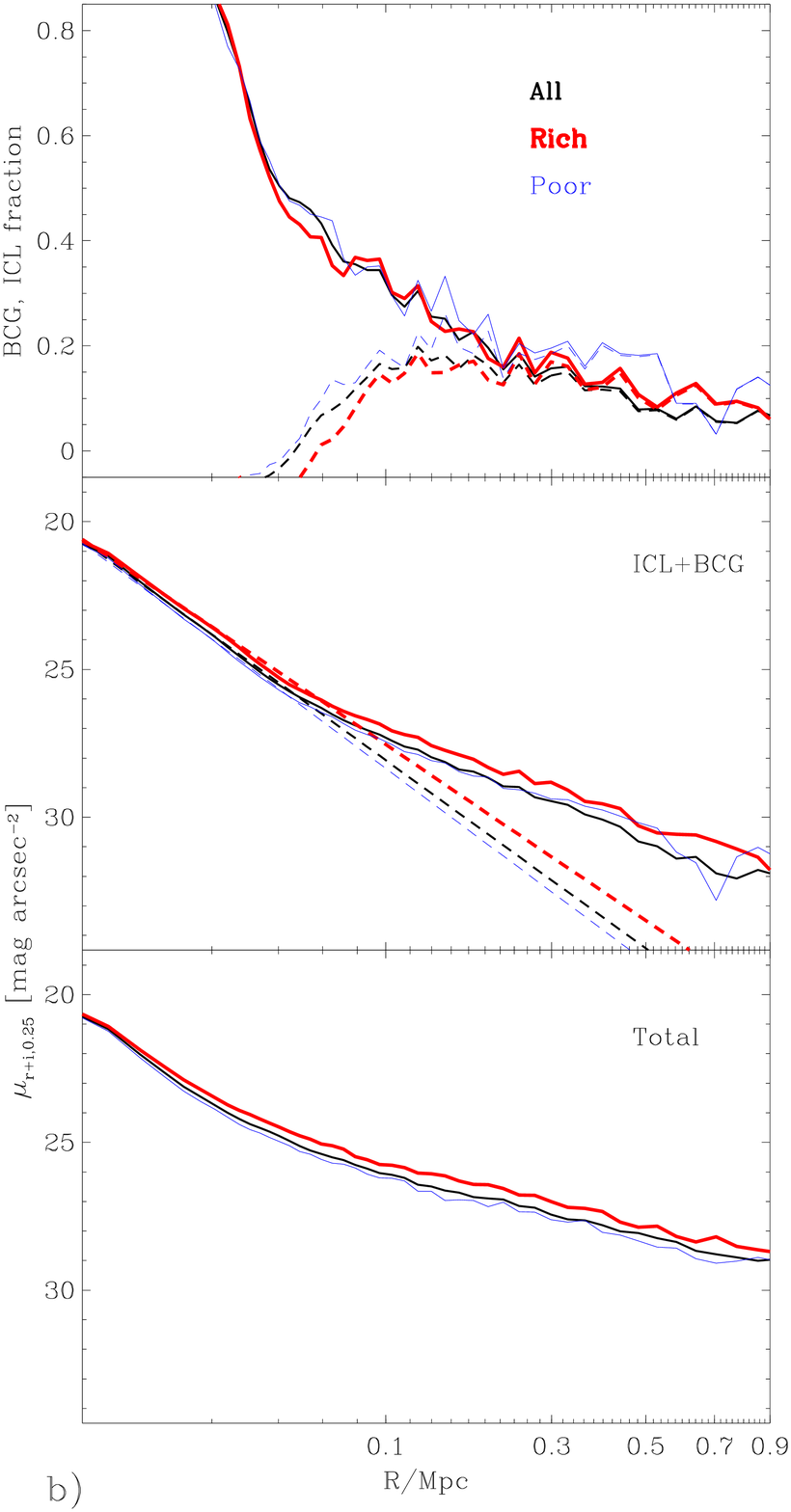}
}\caption{Comparison between different subsamples. Black lines represent the main sample, here
shown as reference. Panel a): clusters with a luminous BCG (red thick lines) vs.
clusters with a faint one (blue thin lines). Panel b): Rich (red thick) vs. poor (blue thin) clusters.
In the three sections of each panel we plot, from the bottom to the
top, the total light SB profile, the ICL+BCG SB profile, with dashed lines representing the 
de Vaucouleurs fitting to the inner data-points, and the ICL+BCG fraction, with dashed lines 
displaying the ICL fraction after subtracting the inner de Vaucouleurs fitting.}\label{prof_comp}
\end{figure*}

In the two graphs of Fig. \ref{prof_comp} we compare the SB profiles and the ICL fractions of
two pairs of subsamples with those of the main sample. In the bottom panels we plot the SB of the
{\it total} light. The panels in the middle display the corrected SB of the {\it ICL+BCG} component; 
the dashed lines represent de Vaucouleurs fits to the inner profiles, derived as explained 
in the previous section. In the top panels we show the local ratio of {\it ICL+BCG} to the 
{\it total} light (solid lines) and the same quantity after subtracting the de Vaucouleurs fit to 
the inner profiles (dashed lines). Black lines are used for the main sample (``All'').

On the left we compare clusters with luminous (``L'') and faint (``F'') BCGs, using red thick lines
and blue thin lines respectively. The SB profiles are very similar outside 100 kpc, but they are offset, 
the ``L'' clusters being brighter than the main sample, and the ``F'' being fainter. 
Except for the central regions, the offset
is similar in the {\it ICL} and in the {\it total} light: ``L'' and ``F'' clusters therefore appear
to have similar relative amount of ICL with respect to the total emission.
However, we note a systematic enhancement of the ICL fraction by a few per cent in the ``L'' 
relative the to ``F'' clusters.
The largest differences are observed in the central regions, where
the emission is dominated by the BCG: the most luminous BCGs have larger effective radii 
($R_e\sim23$ kpc) than the mean ($R_e\sim19$ kpc), whereas faint BCGs have smaller effective
radii ($R_e\sim11$ kpc). The difference in luminosity of the de Vaucouleurs fits to the two
classes is a factor of 2.3.

The comparison between clusters of different richness, as determined from the number of
red-sequence galaxies, is illustrated by the graphs on the right. Here red thick lines represent rich
clusters ($N_{\mathrm{gal}}>22$), while blue thin lines are used for poor clusters 
($N_{\mathrm{gal}}<17$). Again we observe a great similarity between the profiles. As expected,
richer clusters are brighter than the mean, and poor clusters are fainter, both in the {\it total}
emission and in the {\it ICL}. Within 700 kpc the ratio of the {\it total} luminosities of the two
classes is 1.8 (1.7 for the {\it ICL}). The effective radius of the central $R^{1/4}$ profile is 
somewhat larger in the rich clusters ($R_e\sim23$ kpc, similar to ``L'' clusters), than in the poor
ones ($R_e\sim16$ kpc, close to the average value, and significantly larger than in ``F'' clusters).
At $R>100$ kpc the rich clusters exhibit significantly higher SBs with respect to the mean,
whereas the poor ones are only slightly fainter. Considering the large uncertainties in the
profile of the poor clusters beyond 400 kpc, the fractions of ICL appear to be fully consistent
between the different richness subsamples, ranging from 20 to 5 per cent approximately over the
radius range 150 to 500 kpc.
\\

\begin{figure}
\centerline{
\includegraphics[width=8.5truecm]{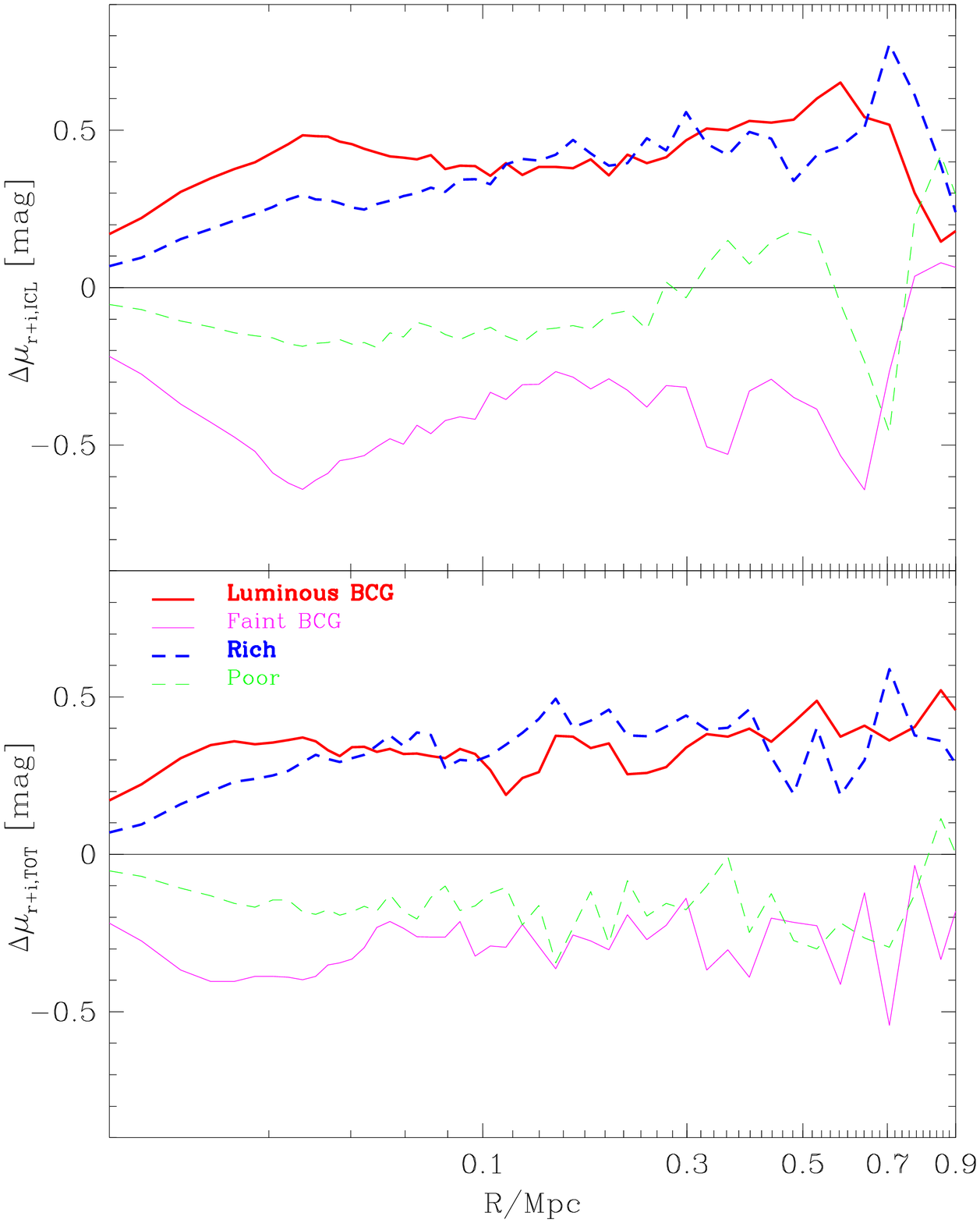}
}
\caption{Comparisons between different subsamples: SB difference between the subsamples and
the main sample. Red thick solid lines are used for clusters hosting a luminous BCG, magenta thin solid
for faint BCG, dashed thick blue for rich clusters, and dashed thin green for poor clusters. 
The bottom panel displays the SB differences as a function of the radius for the {\it total} 
light, the top panel those for the corrected {\it ICL+BCG} component alone.}\label{deltamag}
\end{figure}

In Fig. \ref{deltamag} we compare the same four subsamples, by plotting the
difference in SB with respect to the mean profile, for the {\it total} light (bottom panel)
and for the {\it ICL+BCG} component (top panel). Different samples are represented with different 
lines as indicated in the legend. In the inner 100 kpc we clearly see that the
largest differences are observed between the ``L'' and ``F'' subsamples. This is not surprising,
since this is the region where the BCG dominates. Although the cluster richness   
correlates with the luminosity of the BCG, clusters in the same richness class can have
different BCG luminosity, thus making the separation between rich and poor clusters relatively
small in the centre. At larger radii the influence of the BCG is smaller, and the total SB is
almost equally affected by the richness parameter and by the BCG luminosity. Nevertheless, the
SB of the ICL appears more strongly suppressed in ``F'' clusters than in the poor ones.
\\

\subsection{Dependence on cluster properties: integral photometry}\label{integral}
We further investigate the dependence of the relative luminosity of the cluster components
on the BCG luminosity and on the richness by analysing the integrated photometry of the
stacked $r+i$ images of a set of smaller subsamples. The main sample is thus divided into
5 subsamples according to $m_{r,0.25}$(BCG), and 5 subsamples in $L_{RG}$. 
The $m_{r,0.25}$(BCG) subsamples comprise roughly 140 clusters each, while those in $L_{RG}$ 
have 170 clusters each in the three lower luminosity classes, and 110 and 60 clusters each in the 
two bins at the highest luminosities (these different numbers derive from the skewness 
of the $L_{RG}$ distribution).
The total luminosity of the red-sequence galaxies $L_{RG}$
is used here as a proxy to the richness instead of the number of red galaxies, because of its
property of being a continuous rather than a discrete variable (see Sec. \ref{subsamples}).
In each subsample and for each cluster component (galaxies, BCG, ICL) we measure the integrated 
flux within 500 kpc, and express this as a fraction of the total. The total flux
is obtained directly from the {\it total} light images. The corrected {\it ICL+BCG} flux is
split into a BCG component and the ICL. The BCG flux is given by the integrated flux within 
the radius out to which the inner de Vaucouleurs' profile is fitted plus the integral of the 
fitting profile extrapolated to the outer boundary, while the remaining corrected {\it ICL+BCG} 
flux is attributed to the ``pure'' ICL. Finally, the galaxy flux is just given by the difference
between the {\it total} and the corrected {\it ICL+BCG} flux.\\

\begin{figure*}
\centerline{
\includegraphics[width=8.5truecm]{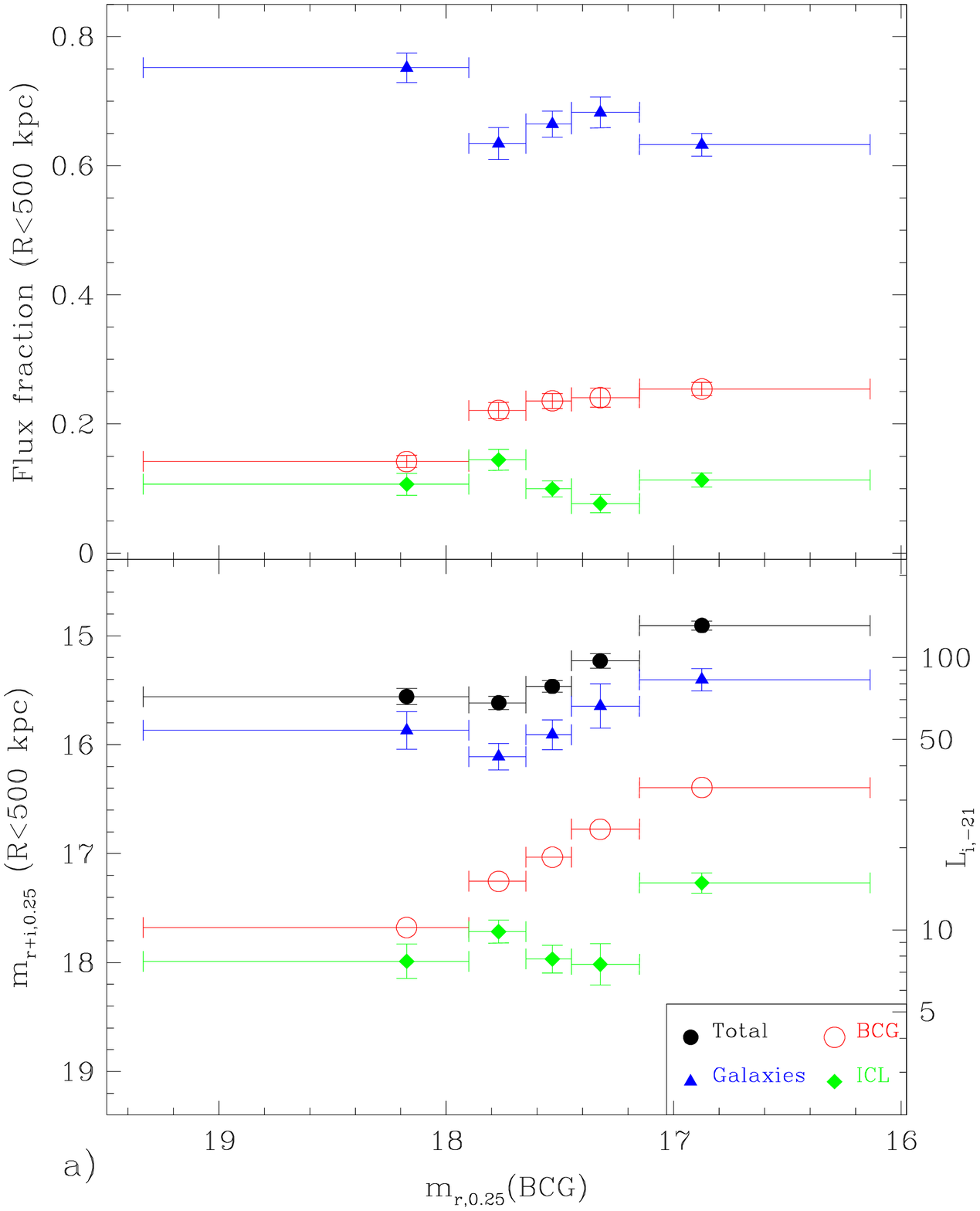}
\includegraphics[width=8.5truecm]{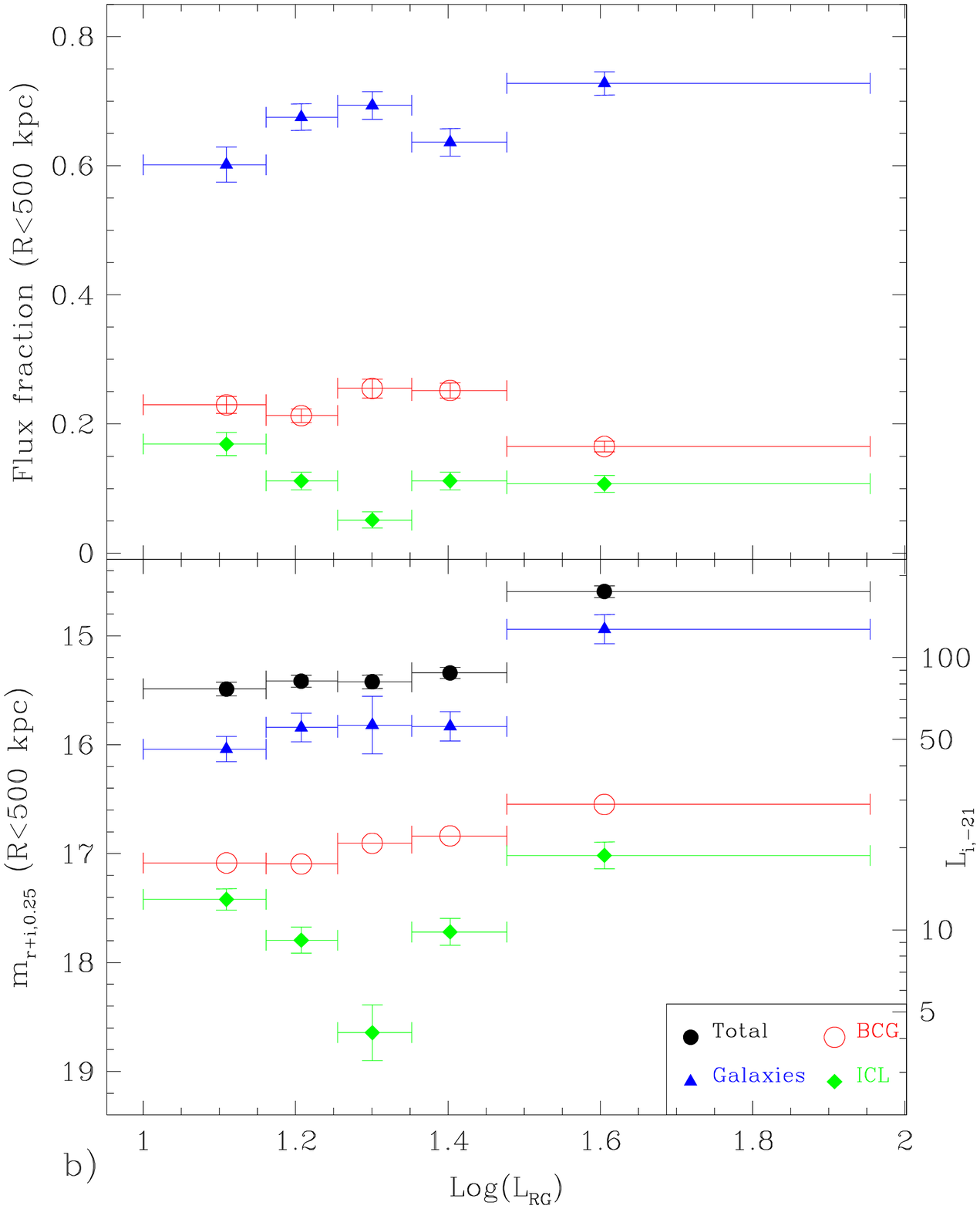}
}
\caption{The dependence of the integrated magnitude within 500 kpc (lower section) 
and of the relative fraction
(upper section) of different cluster components on the luminosity of the BCG
(panel a)) and on the total luminosity of the red-sequence galaxies
(panel b)). BCG luminosity is expressed as $z=0.25$ observer frame
$r$ mag, while the total luminosity of the red-sequence galaxies is given
in units of luminosity corresponding to $-21$ absolute $i$-band mag.
The integrated luminosities are expressed in m$_{(r+i),0.25}$ and in units of $-21$ absolute 
$i$-band mag on the left side and right side axis respectively.
Different symbols and colours are used for the different components as indicated
in the legend box. Horizontal error bars show the range of luminosity encompassed
in each of the four bins, whereas the vertical error bars display the uncertainty
due to the background fluctuations.}\label{intphot_500}
\end{figure*}

The same analysis for the complete main sample yields a ratio galaxies:BCG:ICL of
67.2:21.9:10.9 (uncertainty about $\pm1.0$). Note the small size of the errors here, which is a 
consequence of our very large sample.\\
The results for the subsamples are reported in Fig. \ref{intphot_500},
as a function of $m_{r,0.25}$(BCG) (left panel) and $L_{RG}$ (right panel). The vertical error
bars on the fluxes and fractions take into account background uncertainties and surface brightness
fluctuations within the apertures; no error on the de Vaucouleurs' fit to the BCG is included. 
We caution that such error bars must be regarded as representing the formal photometric errors 
in the stacked images, and do not reflect a fully realistic estimate of the total uncertainties.
These can be inferred approximately from the scatter of values from our 5 independent subsamples
around any smooth trend. The horizontal error bars cover the range of luminosity included in each 
subsample, while the point is plotted at the average value.

Starting with the flux enclosed within 500 kpc as a function of $m_{r,0.25}$(BCG)
(Fig. \ref{intphot_500}, left panel), we note that the luminosity of the galaxy component displays
a weak correlation with the luminosity of the BCG, whereas only the clusters with the most
luminous BCGs have significantly higher ICL luminosity.
The fraction of light provided by the BCG increases from
15 to 25 per cent from the lowest bin to the highest ones, whereas the fraction contributed
by galaxies decreases from 75 to 63 per cent. The ICL percentage, instead, is almost constant around
10 per cent.

As a function of the luminosity on the red sequence (Fig. \ref{intphot_500}, right panel),
we observe weak trends in the total luminosity and in the galaxy and BCG emission. Although the
brightening of $\sim1.1$ mag in the galaxy component is roughly consistent with the 
increase of 0.5 dex in $L_{RG}$ over our 5 bins, a very weak correlation is seen in the four lowest
bins. The richest clusters display substantially higher total and galaxy luminosities. 
Though shallow, a clear correlation between
BCG luminosity and richness is present. The richest clusters appear also to have the most
ICL. Focusing on the relative fractions, we observe no significant trend in 
all components in the 4 lowest bins: the
BCG component represents $\sim23$ per cent of the total luminosity, the galaxies $\sim67$ per cent
and the ICL $\sim10$ per cent. In the richest clusters the contribution of galaxies
grows to 73 per cent, that of the BCG decreases to 16 per cent, while the ICL is responsible for
$\sim11$ per cent of the total flux, as in the other bins.
\\

In both these plots it is consistent to interpret all variations about the mean ICL fraction
of 10.9 per cent just as due to sampling error, that we can estimate around 3--5 per cent.

\section{Systematic uncertainties}\label{systematics}
The results presented in the previous sections are affected by systematic uncertainties that arise
from the methods adopted to estimate the background level and the contamination from galaxy light 
that fails to be masked.
In this section we address these issues by analysing the origins and quantifying the possible
amount of such systematic errors.
\subsection{Background subtraction}\label{systematics_bkg}
As explained above, the background subtraction is based on fitting an NFW profile plus a background
constant to our
raw SB profiles. From previous studies \citep[e.g.][]{NFW_CNOC} we know that the NFW profile fits 
well the mean number density profile for galaxies in clusters. Thus our method appears rigorously
justified as far as the galaxy light is concerned, provided one excludes the cuspy profile of the 
BCG. Given that the galaxy component is dominant, especially at large 
clustercentric distances, the NFW approximation should hold for the {\it total} light profile too.
On the other hand, there is no reason {\it a priori} to expect the ICL to follow any particular
fitting function. Nevertheless, we find that the NFW profile is a reasonably good approximation to
our ICL profile (with an average chi--squared per degree of freedom of $\sim$1.5). Since we are
interested only in a smooth and physically reasonable extrapolation of the SB profile and given 
that the extrapolation required is tiny (the last measured points are just 31--31.5 mag 
arcsec$^{-2}$ above the background), the use of the NFW law appears justified for the ICL as well.\\
Note that the background uncertainties used in the previous section just take statistical 
uncertainties in the fitted background level into account. Different choices for the background
subtraction strategy can yield results differing by up to a few per cent in the integrated
fluxes. Considering the SB profiles, the influence of any reasonable systematic shift of the
background level is negligible for all the points within 500 kpc.
The use of extended image stripes from continuous SDSS scans
would, in principle, provide sufficient coverage to directly estimate the background level at large 
clustercentric distances ($\gtrsim 2~R_{200}$). We will test this possibility in future work.\\
\subsection{Corrections for mask incompleteness}\label{systematics_mask}
In Sec. \ref{corrections} we have computed the fraction of galaxy light that escapes our masking
algorithm, based on the observed surface brightness profiles of galaxies of different luminosities,
and assuming that the number of galaxies as a function of absolute magnitude is well represented
by the luminosity function (LF) of the Coma cluster \citep[][]{mobasher_lf}. 
This is the best studied LF in a rich, massive, regular galaxy cluster, and extends to quite a
faint limit $M_R=-16$; thus our choice is justified. However, the LF of a single rich, massive,
regular cluster at the present epoch may not be representative of the broad range of luminosities
covered by our sample at redshift 0.25 (corresponding to $\sim$20 per cent of the cosmic time).
Therefore, it is worth investigating how the fraction of light that is missed by our masks
changes if a different LF
is used and trying to constrain the possible LF with the available photometric data.

First, using images simulated as described in Sec. \ref{corrections}, we evaluate the fraction of
unmasked light as a function of the absolute $r$ magnitude
of a galaxy and analyse which galaxies are the main contributors of unmasked light. We estimate
its total relative amount in a range of Schechter's function parameters and infer the resulting
ICL fractions.

The fraction $f$ of unmasked light for galaxies
of different absolute $r$ magnitude is reported in Table \ref{cont_tab}.
\begin{table}
\begin{minipage}{8.truecm}
\caption{Fraction of unmasked light $F$ from galaxies of different M$_r$.}\label{cont_tab}
\begin{tabular}{crrrrrr}
\hline
M$_r$&-22.0&-21.0&-20.0&-19.0&-18.0&-17.0\\
$f$ (per cent)& 2.74& 5.02& 8.83&18.21&67.40&99.67\\
\hline
\end{tabular}
\end{minipage}
\end{table}\\ 
Values are close to 0 for bright galaxies.
They smoothly increase up to roughly 0.1 for galaxies of $-20$, and then there is a significant upturn 
at $-19$, that leads to most of the light of faint galaxies being missed by our masks. By integrating
over the entire LF\footnote{We arbitrarily truncate all the LFs at M$_r>-14.0$, in order to make LFs
with $\alpha\leq-1$ integrable. Galaxies fainter than this limit contribute less than 0.1 per cent
of the total luminosity in the range of parameters explored.} we are then able to derive
the relative galaxy luminosity that contributes to the diffuse component. 
\begin{figure}
\includegraphics[width=8.5truecm]{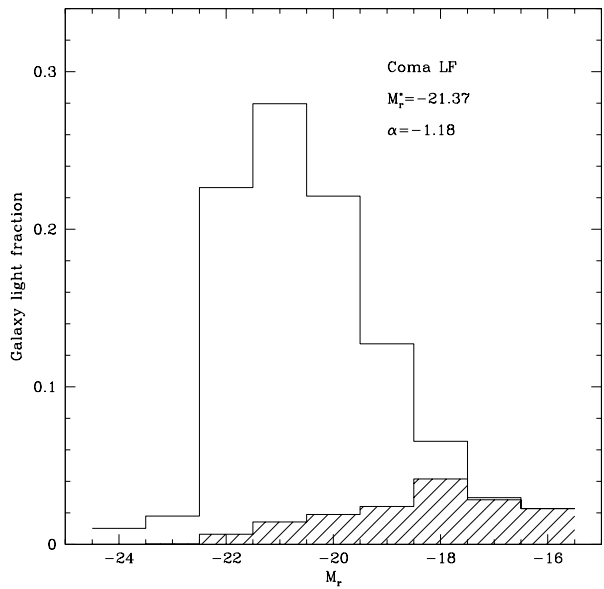}
\caption{The fraction of galaxy light contributed by galaxies of different luminosities (empty
histograms), and missed by our masking algorithm (shaded histograms) adopting the 
Schechter fit to the Coma LF.
}\label{histo_cont}
\end{figure}
In  Fig. \ref{histo_cont} the unshaded histogram shows the relative
contribution of galaxy light in different bins of absolute $r$-band magnitude, according to the 
Coma LF, while the shaded histogram represents the fraction of unmasked light. The first (last) 
bin includes also the contribution from all the galaxies brighter (fainter) than the nominal value. 
While the distribution of galaxy light peaks around the characteristic magnitude $M^*$ of the 
luminosity function, we see that most of the unmasked light comes from galaxies that are at least
2.5 mag fainter than $M^*$.

Dimming $M^*$ increases the contribution of unmasked light from
near-$M^*$ galaxies, while steepening the faint-end increases the contribution from faint
galaxies. Both variations increase the total fraction of unmasked light.
\begin{figure}
\includegraphics[width=8.5truecm]{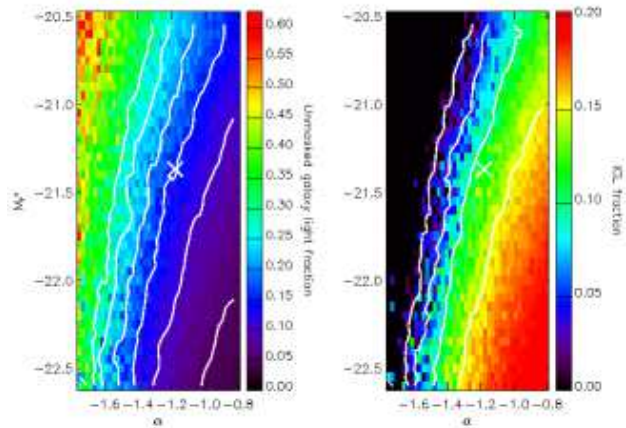}
\caption{Fractions of unmasked galaxy light (left panel) and correspondent ICL estimated fraction
(right panels) as a function of the assumed LF parameters, coded in colour-scale levels according to
the legend bars. Contours show curves of equal fractions in 0.05 intervals, from 0.05 to 0.25
for the unmasked fraction, and from 0 to 0.15 for the ICL. The cross represents the location of
the Coma LF.}\label{cont_map}
\end{figure}
These effects are illustrated and quantified in the left panel of Fig. \ref{cont_map}. For a whole
range of ($\alpha$,$M^*$), the unmasked light fraction is coded
by colour-scale levels, as indicated. Lines display the curves of constant unmasked fraction,
from 0.05 to 0.25 in steps of 0.05, and the cross corresponds to the Coma LF. 
Taking this point as a reference, we see that steepening the faint-end by 0.2 can dramatically
increase the unmasked fraction to $\sim$25 per cent, whereas dimming $M^*$ by 0.4 mag is required
to get to 20 per cent. Going to brighter $M^*$ and less steep slopes, the gradient of unmasked 
fraction $f(\alpha,M^*)$ becomes shallower, so that a fraction of 10 per cent requires brightening
$M^*$ by almost 1 mag, or making the slope shallower by 0.15.

These fractions can be translated into estimates of ICL within 500 kpc, based on the {\it diffuse}
light we have measured, as shown in the right panel of Fig. \ref{cont_map}. Similarly to the 
previous plot, estimated ICL fractions are represented as gray scale intensities in the
($\alpha$,$M^*$) plane. Lines are curves
with the same ICL fraction, from 0 to 0.15 in 0.05 step and the cross corresponds to the Coma LF.
The trend is opposite to the previous plot, brighter $M^*$ and less steep slopes implying higher
ICL fraction. The range of variation is between 0 and 20 per cent. A very interesting feature of
this plot is the presence of a ``zone of avoidance'' to the left of the 0 level contour, where the
the amount of unmasked light would be larger than the {\it diffuse} light. This excludes all the
LFs with very steep faint-end\footnote{This argument rigorously applies only to a Schechter LF. 
However, the excess of dwarf ellipticals fainter than $M_R=-15$ found in Virgo 
\citep[e.g.][]{trentham_hodgkin02,sabatini_etal03} are unlikely to produce additional contamination
larger than 1 per cent.} $\alpha\lesssim -1.5$. Applying the same argument to the local fraction
of {\it diffuse} light shown in Fig. \ref{profiles}, we conclude that the fraction of unmasked
light cannot be larger than 20 per cent. In turn, this implies {\it i)} that the fraction of ICL integrated
within 500 kpc must be at least 6.5 per cent, and {\it ii)} that a faint-end steeper than $\alpha=-1.35$
is inconsistent with our data, unless we assume $M^*$ very different from the reference value for 
Coma.\\
As an additional piece of evidence against very steep faint-end LF,
we note that the colours of the ICL shown in Fig. \ref{profiles} d) are the same as or marginally
redder than those of the total light. If the ICL was dominated by dwarf galaxies, any
reasonable colour-magnitude relation would imply bluer colours than those of the total light.

The natural upper limit to the ICL fraction is given by the fraction of {\it diffuse} light, 
that is 21 per cent, which is obtained in the limit of no unmasked light.
On the other hand, if we adopt the ``brightest'' LF in literature, namely the one by \cite{goto_lf} 
($\alpha=-0.85$, $M_r^*=-22.21$) based on 204 SDSS nearby clusters, we obtain an ICL fraction of 
17 per cent ($f=5$ per cent). This LF has a very bright $M_r^*$ and a very shallow faint-end in 
comparison to all other measurements in literature.\\
Although we cannot go very deep and investigate the faint-end,
we have measured the composite LF of our clusters (excluding the BCGs) by using the SDSS 
photometric catalogues and applying statistical subtractions. Our LF is complete to $r=22$ 
($M_r=-18.4$) and clearly indicates that $M^*$ is the same as in Coma within a few 0.01 mag,
sufficient to exclude the LF of \cite{goto_lf} as a good description of our clusters and to limit
the possible choice of LF parameters.\\
Adding the fact that, in the light of most of the LF studies in literature,
$\alpha>-1$ seems implausible, we arrive at a tightly constrained region in the space of
possible parameters: $M_r^*=-21.37\pm0.1$, $-1.35<\alpha<-1$. 
This translates into an uncertainty of $\pm5$ per cent in the estimated fraction of ICL.
\section{Discussion}\label{discussion} 
In previous sections and particularly in Fig. \ref{profiles}, we have presented surface
photometry from the stacking of 683 clusters of galaxies (our main sample). We have been able to
measure SB as deep as $\mu_{r,0.25}\sim 32$ mag arcsec$^{-2}$ for the ICL light and
$\mu_{r,0.25}\sim 29.0$ for the total light, out to
700 kpc from the BCG. Such SBs translate into rest-frame $g$-band SBs which are roughly 
1 mag arcsec$^{-2}$
brighter, as can be easily seen considering the $(1+z)^4$ cosmological dimming and the fact that 
the $r$ band at $z=0.25$ is centred at $\lambda_{\mathrm{eff}}'=
\lambda_{\mathrm{eff}}/(1+z)\sim 5000$ \AA.

Looking at the profiles reported in Fig. \ref{profiles} we see that: {\it i)} the inner $\sim50$
kpc are well reproduced by an $R^{1/4}$ law; {\it ii)} a significant excess of diffuse light
with respect to the inner $R^{1/4}$ law is clearly seen beyond 100 kpc out to 700 kpc;
{\it iii)} the SB of the ICL decreases faster than the total SB of the cluster, i.e. the ICL
is more centrally concentrated than the light of the cluster as a whole.
The inner de Vaucouleurs component is comparable in size with the BCGs observed by 
\cite{schombertBCG}. In fact, the average profile obtained from the stacking of our main sample has 
an effective radius $R_e\simeq 19$ kpc, which roughly corresponds to the median value of $R_e$
in his sample.

SB excesses with respect to an $R^{1/4}$ law at large radii ($\gtrsim100$ kpc)
from the BCG have recently been observed by \cite{gonzalez+04} in a sample of 24 clusters at 
$0.03<z<0.13$. They represent these excesses as a second $R^{1/4}$ component with larger $R_e$.
Parametrising the SB excesses in our data in terms of $R^{1/4}$ law between 150 and 500 kpc, 
we find an effective radius $R_e\sim$ 250--300 kpc, consistent with the range of outer $R_e$ 
measured by \cite{gonzalez+04}, although their distribution is peaked at a somewhat lower value, 
around 100 kpc. However, the ratio of the $R_e$ of our two components is 0.06--0.07, fully consistent 
with their results.
The ICL is significantly more concentrated than the total light. Uncertainties or bias deriving
from background subtraction and from the estimate of contamination are very unlikely to change this
result significantly.

In contrast to our own results and those of \cite{gonzalez+04}, \cite{feldmeier+04c} have recently 
claimed that the intracluster
star density, as measured from planetary nebulae, does not change as a function of radius
or projected galaxy density in the Virgo cluster. This may partly be explained by the
young dynamical status of this cluster. However, large field-to-field variations in their
estimated SB and their sparse sampling of the cluster region prevent us from drawing firm 
conclusions. In particular, it is noticeable that the fields at the largest distances from M87 
are actually close to M49, which is known to be associated with a major sub-cluster: it is no 
surprise, therefore, that the estimated SB in those fields is much higher than expected from any 
SB--radius relation.

Numerical simulations published to date make a variety of different 
predictions for the slope of the outer de Vaucouleurs component, ranging from $\sim400$ kpc 
\citep{willman+04} to 70--100 kpc \citep{sommerlarsen+04}. Our results seem to favour the
models with larger $R_e$.
It is interesting to note that \cite{murante+04} produce 
steeper profiles for the ICL than for the galaxy component in their simulated clusters,
and hints of similar behaviour are visible in the simulation of \cite{willman+04} as well.

Is the measured SB excess contributed by genuine intracluster stars, that is stars orbiting freely 
in the cluster potential rather than bound to individual galaxies?
There are several indications that this is actually the case.
The change in slope of the $ICL+BCG$ profile and the change in its colour gradient at $R\sim70$ kpc
suggests strongly that the BCG and ICL components can be considered as distinct stellar populations
with different assembly histories.
We note that the diffuse light extends continuously well beyond the radius 
$R\sim 300$ kpc at which the enclosed stellar mass (as traced by the light) begins to be 
dominated by galaxies other than the BCG.
At larger clustercentric distances the dynamics of the diffuse stellar population must
be dominated by the cluster potential, rather than that of the BCG.
The fact that we do not see any discontinuity in the diffuse light profile nor
any colour gradient going from 100 kpc
to the outer regions, lends support to the idea that the stars contributing the
SB excess in the inner regions have similar dynamical properties to those at larger
distances, and so also orbit freely in the cluster potential.

Based on our analysis of the correlation between the diffuse light and the 
galaxy distribution, we can exclude the possibility that all or most of the diffuse light is 
physically linked to individual non-central galaxies, at least at projected radii below 300 kpc.
This conclusion is reinforced by the different concentration of the two components.

Surface brightness
excesses spatially associated with bright galaxies on scales up to 40 kpc
contribute significantly to the ICL. 
For example, at clustercentric distances of 400--750 kpc, SB excesses surrounding galaxies on this
scale sum up to $\gtrsim80$ per cent of our total measured ICL. From our 
analysis alone we cannot argue about the origin of this excess. However, 
observations of individual clusters reported by several authors 
\citep[e.g.][]{trentham_mobasher98,gregg_west98,calcaneo+00,feldmeier+02,feldmeier+04},
suggest that tidal structures like plumes and arcs may be good candidates for some of it. In fact,  
smooth bound low-surface brightness haloes around individual galaxies appear unlikely to account for
all the light excess,
given the large spatial scale, which is larger than the typical optical size of cluster galaxies
\citep[e.g.][]{VCC} and comparable with or larger than the expected tidal radii
of galaxies in clusters \citep{merritt84}.\\ 

Further support for our identification of the diffuse light as a distinct component comes from our
analysis of isophotal shapes for clusters where the BCG is substantially flattened: in Fig.
\ref{ellipticity} we have shown that the ICL is significantly more flattened both than the BCG 
``core'' itself and than the galaxy distribution.
Examples of flattening of the BCG's outer halo with respect to its inner parts have been known
since the late 1970s \citep{dressler79,porter+91}, as an
association with a similar flattening of the galaxy distribution 
\citep[e.g.][]{binggeli82}. Our ellipticities and the corresponding radial dependences are 
completely consistent with those of \cite{gonzalez+04}. Moreover, extending the observed radial
range well beyond 100--200 kpc, 
we can present evidence for an asymptotic value for this ellipticity, which is only
suggested by their data, although predicted by the two-component de Vaucouleurs
models which they fit to the BCG+ICL surface brightness distribution. The fact that this
asymptotic ellipticity is first reached where the slope of the SB of the diffuse light flattens
lends further support to the hypothesis of a distinct second component responsible for the outer
profile. In addition to this, it is intriguing that the change in slope and 
ellipticity occurs where the galaxy component begins to dominate the total SB, apparently 
establishing a link between the galaxies and the ``true'' ICL.\\

During the last decade many attempts have been made to assess the total amount of ICL and its
contribution to the total cluster light. Current estimates based on different methods
range from less than 10 per cent for poor groups of galaxies \citep{feldmeier_ciardullo+04}
to $\lesssim 20$ per cent for non-cD clusters \citep{feldmeier+04}, to 20--40 per cent for cD
clusters \citep{schombert88,feldmeier+02} and up to $\sim 50$ per cent for Coma
 (\citealp[$R\lesssim 500$ kpc,][]{bernstein+95}, but at $<25$ per cent according to \citealp{melnick_white_hoessel77}).
The results presented in Sec. \ref{integral} indicate that in the mean the ICL contributes 
$10.9\pm1$ per cent of the flux within 500 kpc, while the de Vaucouleurs component of the
BCG contributes $21.9\pm1$ per cent. We warn that the uncertainties reflect the 
measurement errors only, and we expect the overall uncertainty (sampling plus systematic) in this 
measure to be about 5 per cent for the ICL and 3 per cent for the BCG. Variations between 
individual clusters are of course likely to be much larger.
Our results thus favour a quite low average ICL fraction, compared to previous estimates.
This conclusion is not particularly biased by the properties of our sample, where intermediate
and low mass clusters dominate: similar fractions are obtained almost independent of cluster richness
and BCG luminosity. This apparent discrepancy points to the problem of how estimates of ICL
are derived with different methods and to the need to obtain reliable cross calibrations.

In our analysis of different subsamples, binned in luminosity of the BCG and in richness,
we found that richer clusters, and those with a more luminous BCG,
have brighter ICL than poor clusters or clusters with a faint BCG.
If we consider the local ICL fractions, however, the variations between
different classes are no more than $\pm 5$ per cent, because the total SB varies roughly in the 
same way as that of the ICL. 
On the other hand, binning the clusters in finer $L$(BCG) and richness classes, we found that
a significantly higher ICL luminosity within 500 kpc is measured only in the richest clusters and
those having the most luminous BCGs, while no trend is observed in the other classes.
As already stated, the fraction of ICL instead is roughly constant, within the uncertainties and
the sample variance.\\
We stress that the quoted luminosities are integrated within fixed metric apertures
in all our subsamples. These apertures correspond to different fractions of the virial radius
in different clusters and this must be taken into account when comparing our results with 
fluxes and fractions computed within the virial radius. In fact, a rough extrapolation of
the growth curves to the total luminosity $L_{200}$
within $R_{200}$ shows that for the poorest clusters $L_{200}$ is roughly 1.7 times the luminosity
within 500 kpc, whereas for the richest clusters $L_{200}$ is roughly 2.5 times this luminosity.
This particularly affects the light fraction contributed by the BCG: although in the fixed 500 kpc
aperture the BCG represents an almost constant fraction of the total light, independent of the
cluster richness, the same fraction within $R_{200}$ would show a decreasing trend with  
richness, as found by \cite{lin_mohr04}.

The analysis of the colour profiles in Fig. \ref{profiles} (d) demonstrates that the ICL colours
are consistent with (in $g-r$) or marginally redder than (in $r-i$) the average colours
of galaxies. This result is compatible with the idea that the ICL originates from stripped stars 
and disrupted galaxies. Given the relatively large uncertainties in
our colour estimates, we cannot test the slightly different predictions obtained by the recent 
N-body+SPH simulations of \cite{murante+04},\cite{willman+04} and \cite{sommerlarsen+04}.
In fact, they all agree in predicting that intracluster stars must have roughly the same colours
and metallicities as the dominant stellar population in galaxies, but while \cite{murante+04} and
\cite{sommerlarsen+04} predict slightly larger ages for the intracluster stars, \cite{willman+04}
argue that the typical intracluster stellar population should be similar to those in intermediate
luminosity galaxies.\\

As a final remark, we note that 
in a scenario where the ICL originates from stripping and galaxy disruption, the 
galaxies that contribute most of the ICL are those plunging into the cluster potential
along nearly radial orbits \citep[e.g.][]{moore+96}, with some of them eventually merging into the BCG.
If there is a significant
anisotropy in the orientation of the orbits, a significant elongation in the shape of the BCG
and of the ICL should be observable. However, because of the shorter orbital and scattering times at
higher densities, the elongation is expected to increase with increasing clustercentric distances,
up to the asymptotic value given by the ``original'' distribution of orbital parameters.
This may explain the outwardly increasing ellipticities of the ICL isophotes which we found 
above (see Fig. \ref{ellipticity}). \\

\section{Conclusions}\label{conclusions}
In this paper we have studied the mean properties of the intracluster optical emission
of 683 clusters of galaxies between $z=0.2$ and 0.3, imaged by the SDSS. Thanks to the high
sensitivity achieved by stacking the imaging data, we have been able to trace the average 
SB profile of the ICL out to 600--700 kpc from the BCG. Measured SB ranges from 27.5 mag 
arcsec$^{-2}$ at 100 kpc to $\sim 32$ mag arcsec$^{-2}$ at 700 kpc in the observed
$r$ band, which corresponds to $\sim 1$ mag arcsec$^{-2}$ brighter rest-frame $g$-band SB.
The ICL fraction depends at most
weakly on global cluster properties, such as BCG luminosity and richness. The ICL is ubiquitous 
in clusters of galaxies, as demonstrated by significant detections in all our subsamples. 

We find that the ICL contributes in the mean 30--40 per cent of the total optical emission at  
around 100 kpc and a decreasing fraction at larger clustercentric distance, down to $<5$ per cent
at 600--700 kpc. By integrating the fluxes of the different components within 500 kpc
we obtain $10.9 \pm 1.0$ per cent for the fraction of light in the ICL and $21.9 \pm 1.0$ per cent 
for the BCG. Taking sampling uncertainties and systematic errors into account, the total errors 
on these fractions are about $\pm 5$ per cent for the ICL and $\pm 3$ per cent for the BCG.
Our measurements of the diffuse light put also an independent constraint on the shape of the
cluster luminosity function: faint-end slopes $\alpha<-1.35$ are rejected as inconsistent.

The higher spatial
concentration of the ICL with respect to the starlight in galaxies, indicates that the 
production mechanism for the ICL is more efficient the deeper one goes into the cluster potential 
well. Comparing different subsamples of clusters, we have observed a significant correlation of the
surface brightness of the ICL with the luminosity of the BCG, as well as with the richness of the 
cluster, suggesting a link between the mechanisms responsible for the growth of the BCG and
for the accumulation of intracluster stars.

The similarity in colours between the ICL and galaxies supports a scenario where intracluster
stars originate in galaxies and are subsequently dispersed in the intracluster space by
dynamical interactions leading to galaxy stripping or disruption. Moreover, the analysis of the
shape (ellipticity) of the ICL with respect to the BCG core suggests that the main mechanism 
acting to create the ICL is the tidal interaction of galaxies with the central cluster potential.
This would also explain the observed link between the amount of ICL and the BCG's luminosity,
the latter being strongly correlated to the depth of the potential well.

Due to the very large sample size and to the unprecedented depth of the present observations,
our results provide the best and statistically most representative measurement of the
intracluster light so far over a wide range of cluster types.
Future extensions of our sample to the entire SDSS area and to nearer clusters, and improvements in
classification algorithms for galaxy clusters will not only improve the sensitivity of our
measurements, but also
make it possible to study the relationship between cluster properties and the ICL in greater detail,
providing new clues and stronger constraints on dynamical processes during the formation and 
evolution of galaxy clusters.\\

\section*{Acknowledgements}
We thank Jim Annis for kindly providing us with the maxBCG catalogue of clusters used in this
work, and Sarah Hansen for making available unpublished results on the photometric determination
of $R_{200}$. Thanks to the referee Neil Trentham for comments that have significantly improved
this paper, to Anthony Gonzalez and Magda Arnaboldi for very stimulating discussions
and the fruitful comparisons of respective results, and to Anna Gallazzi for her precious help.\\
This paper is dedicated to Jatush Sheth.

Funding for the creation and distribution of the SDSS Archive has been provided by the Alfred 
P. Sloan Foundation, the Participating Institutions, the National Aeronautics and Space 
Administration, the National Science Foundation, the U.S. Department of Energy, the Japanese 
Monbukagakusho, and the Max Planck Society. The SDSS Web site is http://www.sdss.org/.\\
The SDSS is managed by the Astrophysical Research Consortium (ARC) for the Participating 
Institutions. The Participating Institutions are The University of Chicago, Fermilab, the 
Institute for Advanced Study, the Japan Participation Group, The Johns Hopkins University, 
the Korean Scientist Group, Los Alamos National Laboratory, the Max-Planck-Institute for 
Astronomy (MPIA), the Max-Planck-Institute for Astrophysics (MPA), New Mexico State University, 
University of Pittsburgh, Princeton University, the United States Naval Observatory, and the 
University of Washington.

\label{lastpage}
\end{document}